\documentclass{aa}

\usepackage[varg]{txfonts}
\usepackage{graphicx}
\bibpunct{(}{)}{;}{a}{}{,} 

\usepackage{xcolor}

\usepackage{color}
\usepackage[colorlinks,linkcolor=blue,citecolor=blue,urlcolor=blue]{hyperref}

\begin{document}

        \title{Ionization of heavy elements and the adiabatic exponent in the solar plasma }

        \author{V.A.~Baturin\inst{\ref{inst1}}
        \and A.V.~Oreshina\inst{\ref{inst1}}
        \and W.~D\"appen\inst{\ref{inst2}}
        \and S.V.~Ayukov\inst{\ref{inst1}}
        \and A.B.~Gorshkov\inst{\ref{inst1}}
        \and V.K.~Gryaznov\inst{\ref{inst3}}
        \and I.L.~Iosilevskiy\inst{\ref{inst4},\ref{inst5}}}

\institute{Sternberg Astronomical Institute, M.V. Lomonosov Moscow State University, 13, Universitetskij pr., 119234, Moscow, Russia \email{avo@sai.msu.ru}\label{inst1}
        \and
        Department of Physics and Astronomy, University of Southern
        California, Los Angeles, CA 90089, USA\label{inst2}
        \and
        Institute of Problems of Chemical Physics RAS, Chernogolovka, Russia\label{inst3}
        \and
        Joint Institute for High Temperatures RAS, Moscow, Russia \label{inst4}
        \and
        Moscow Institute of Physics and Technology, Dolgoprudnyi,
        Russia \label{inst5}
 }

\date{Received 26 July 2021 /
      Accepted 31 January 2022}

\abstract{}{}{}{}{}
 \abstract
 {The adiabatic exponent  $\Gamma_1$ is studied as a thermodynamic quantity in the partially ionized plasma of the solar convection zone.}
 {The aim of this study is to understand the impact of heavy elements on the $\Gamma_1$ profile. We calculated $\Gamma_1$ with the SAHA-S equation of state for different chemical compositions of plasma, and we analyzed contributions of individual elements to $\Gamma_1$. We attempted to determine the mass fractions of the heavy elements using our analysis of the $\Gamma_1$ profile.}
 {We studied the decrease in $\Gamma_1$ due to the ionization of heavy elements in comparison with the value obtained for a pure hydrogen-helium plasma. These types of differences are denoted as ``$Z$ contributions'', and we analyzed them for eight elements (C, N, O, Ne, Mg, S, Si, and Fe) as well as for a mixture of elements corresponding to the solar chemical composition. The contributions of the heavy elements are studied on an adiabat  in the lower part of the convection zone, where the influence of hydrogen and helium to the $Z$ contribution is minimal. The $Z$-contribution profiles are unique for each chemical element. We compared linear combinations of individual $Z$ contributions with the exact $Z$ contribution.  Applying a least-squares technique to the decomposition of the full $Z$ contribution to a basis of individual-element contributions, we obtained the mass fractions of the heavy elements.}
 {The $Z$ contribution of heavy elements can be described by a linear combination of individual-element $Z$ contributions with a high level of accuracy of $5\times 10^{-6}$ . The inverse problem of estimating the mass fractions of heavy elements from a given $\Gamma_1$ profile was considered for the example of solar-type mixtures. In ideal numerical simulations, the mass fractions of the most abundant elements could be determined with a relative accuracy better than a few tenths of a percent. In the presence of random or systematic errors in the $\Gamma_1$ profile, abundance estimations become remarkably less accurate, especially due to unknown features of the equations of state. If the amplitude of the errors does not exceed $10^{-4}$, we can expect a determination of at least the oxygen abundance with a relative error of about 10\%. Otherwise, the results of the method would not be reliable.
 }
 {}

      
\titlerunning{Ionization of heavy elements and the adiabatic exponent}

   \maketitle


\section{Introduction}

Properties of the adiabatic exponent $\Gamma_1$ are studied for conditions of the solar interior. The thermodynamic quantity $\Gamma_1$ describes the variation of pressure $P$ under adiabatic compression. It is determined by the formula $\Gamma_1={\left. {\partial \ln P}/{\partial \ln \rho }\; \right|}_S$, where $\rho $ is the density and $S$ is the entropy.

It is commonly known that in a mono-atomic ideal gas ${{\Gamma }_{1}}={5}/{3}\;$. In the present paper, we are mostly interested in the deviation from this value due to the ionization of chemical elements. It is well known that a region of ionization generally contributes to a lowering of $\Gamma_1$ from its mono-atomic value. There are several analytical expressions that illustrate the effect of the hydrogen ionization on $\Gamma_1$  \citep{Cox_Giuli_1968, Hansen_2004}. Also, further examples of analytic expressions are given in Appendix~\ref{Appendix_Ionization}. The typical result for a one-stage ionization is a bell-shaped decrease in $\Gamma_1$  as a function of the ionization degree.

Interpretation of this decrease in $\Gamma_1$ is analogous to inferences from the decrease in intensity in an absorption spectral line. While in the latter the line position indicates the transition energy and the line depth the mass fraction of the chemical element, in the case of $\Gamma_1$ the position of the decrease indicates an ionization zone of a given chemical element and the depth of the decrease is proportional to the abundance of that element.

In the deeper layers of the solar convection zone (CZ), the stratification is nearly adiabatic. In that part of the Sun, the $\Gamma_1$ profile can be observationally inferred with high accuracy, since uncertainties from convection matter little. Thus, the observational $\Gamma_1$ can be used to study theoretical models for the ionization of various elements in a wide range of temperatures and densities.

The computation of thermodynamic values for a mixture of partially ionized elements requires an advanced equation of state, which -- besides ionization -- takes into account complex physical effects: charged particle interaction, electron degeneration, and relativistic effects. These elaborated methods lead to significant computational costs, which is why interpolations in pre-computed tabulated data are often used. Some of the problems associated with the computation and interpolation of $\Gamma_1$ in modern equations of state are described in \citep{Hansen_2004, Baturin_2017, Baturin_2019} and in Appendix~\ref{Appendix_Ionization}.  
In the present paper, we {\bf } used the data of the SAHA-S equation of state \citep{Gryaznov_2006, Gryaznov_2013} (see also the SAHA-S
web-site\footnote{\href{http://crydee.sai.msu.ru/SAHA-S_EOS}{http://crydee.sai.msu.ru/SAHA-S\underline{\hphantom{a}}EOS}}). 
This equation of state was chosen on the matter of high internal precision and consistency of thermodynamic functions. 
This is a requirement for an analysis of rather small effects associated with heavy elements.

Investigations of the effect of ionization on adiabatic exponents started in the 1980s. \citet{Dappen_Gough_1984} showed that the sound speed and $\Gamma_1$ significantly decrease in the second helium-ionization region of the solar convection zone. This ionization of helium has captured attention, because its effect on $\Gamma_1$ is strong and can be used for a helioseismological estimation of the helium abundance $Y$. \citet{Kosovichev_1992} noticed that such an estimation of $Y$ can be influenced by the $\Gamma_1$ variation due to heavy elements. The authors examined several different theories of the ionization of heavy elements and concluded that helioseismological inversions are sensitive to this description. \citet{Dappen_1993} also studied the influence of heavy elements on $\Gamma_1,$ with an emphasis on the impact of excited states in compound species.

Since the stellar plasma contains a mixture of heavy elements, the relative abundances of the elements in the mixture are of interest. \citet{Elliott_1996} studied the possibility of estimating both total and individual contributions based on the $\Gamma_1$ profile. He used derivatives $\partial {{\Gamma }_{1}}/\partial {{f}_{i}}$ with respect to parameters ${{f}_{i}}$ that describe the heavy-element mass fractions ${{Z}_{i}}$. The derivatives are a linear combination of all element contributions. From a physical viewpoint, such derivatives are similar to the contributions to $\Gamma_1$ from element mass fractions used in our work.

Element contributions were analyzed in the paper by \citet{Gong_2001}. Individual contributions of heavy elements were computed  by subtracting $\Gamma_1$~profiles obtained in different approximate solar models. The models were computed for special mixtures, in which the 2\% of heavy-element mass was assigned to only one element (C, N, O, or Ne). Such models were compared to models without heavy elements. The element contributions were computed using several different equations of state (OPAL and MHD, among others). The authors showed that $\Gamma_1$ decreases in the lower part of the convection zone in the lg T=[5.5, 6.5] temperature range, where electrons are detached from the inner K envelope of  ions. At that time, the authors found that the decrease is due entirely to the heavy-element contribution and is almost independent of the chosen equation of state. In our work, we compared $\Gamma_1$ for different chemical compositions at fixed $({{T}_{j}},{{\rho }_{j}})$ points  while keeping the same equation of state. This allowed us to focus solely on the thermodynamics of the $\Gamma_1$ decrease.

A helioseismological determination of the heavy element mass fraction $Z$ in the convection zone has been attempted by several authors.  The paper by \citet{Baturin_2000} examined the thermodynamic effect of heavy elements on the $\Gamma_1$ profile. The possibility of a helioseismological estimation of $Z$ based on $\Gamma_1$ was considered in the works by \citet{Baturin_2000} and \citet{Lin_Dappen_2005}. \citet{Antia_Basu_2006} and \citet{Lin_2007} obtained $Z$ = 0.0172. However, \citet{Vorontsov_2013, Vorontsov_2014} estimated  Z = 0.008–0.013. The low $Z$ value is confirmed by helioseismic inversions in \citet{Buldgen_2017}. Their results are close to recent spectroscopic data \citep{Asplund_2005, Asplund_2009, Asplund_2021}.  Inversions by \citet{Vorontsov_2013, Vorontsov_2014} indicated a slightly higher $\Gamma_1$ on the Sun compared to the standard model. This result can be matched by a modern equation of state with a low $Z$.

Our present paper continues the work by \citet{Baturin_2010} and \citet{Baturin_2013}, where a method for the individual $Z$ contributions was developed together with an application and improvement of the SAHA-S equation of state. \citet{Baturin_2010} used a SAHA-S version with four heavy elements (C, N, O, and Ne). Later computations by \citet{Baturin_2013} were based on a version with an extended chemical composition (8 elements, including Fe, Si, Mg, and S) and a detailed physics of the excited states of all compound species.

The main purpose of the paper is to study the physical and mathematical foundations that enter the $\Gamma_1$ profile in the relevant region of the Sun, which lies in the solar convection zone. The role of the $\Gamma_1$ profile for seismic abundance determinations is indisputable and well established.
        
The physical principles studied in the paper are based on the sequential ionization of each species of a mixture of heavy elements as well as the computation of the sum of all individual effects. It turns out that there are two main results concerning the additivity of the effects in the resulting profile of $\Gamma_1$. First, we have additivity of the contributions from different heavy elements, and second, we have additivity of contributions from the sequential ionization stages of a particular element.
        
Mathematically, the function of $\Gamma_1$ localized in a piece of the solar convection zone is decomposed with respect to a set of basis vectors. The basis vectors, belonging to the effect of each heavy element on $\Gamma_1$, span a linear space. The idea is that a given $\Gamma_1$ profile will reveal the abundances of the heavy elements through the coefficients of the basis vectors. We want to emphasize that the basis functions are determined by thermodynamics, in particular the atomic structure of chemical elements. They depend of course on the details of the underlying equation of state. However, our analysis is dedicated to the process of adding the various effects. Surely, any error in the basic atomic and thermodynamic theory would lead to an overall uncertainty of the final result. However, the analysis of such errors will have to be done in the future, when the method is applied to real helioseismic data. In the present work, we deliberately restricted ourselves to the mechanism of addition of the various effects.
        
The element-induced dips of $\Gamma_1$ in ionization regions are well known and have been extensively studied before. To isolate the relevant effects, in our paper we introduce so-called $Z$ contributions instead of the $\Gamma_1$ profile itself. Using $Z$ contributions improves the sensitivity to abundances, but it reduces the domain of applicability to a "space" around the exact ideal $\Gamma_1$ profile. We computed this ideal solution and demonstrated that there is a neighborhood around it where the method works.
        
In this work, we did not touch on the question of determining the heavy-element abundance in the Sun. The reason is that there are too many factors precluding such an application at the moment: (1) the absence of a good $\Gamma_1$ from inversions, (2) the unknown entropy in the convection zone, (3) the presently used equations of state might still not be sufficiently close to reality, and so on. Therefore, the present paper is somewhat preliminary. We chose solar conditions merely in order to make our study as realistic as possible, not to imply that the method can be applied to the Sun immediately.

The aim of our work is (i) to estimate the effect of the individual heavy elements on the $\Gamma_1$ profile and (ii) to find a solution to the inverse problem; that is, in determining the mass fractions ${{Z}_{i}}$  out of the overall contribution of the heavy elements to $\Gamma_1$.

We give results of numerical computations as well as analytical formulae for a better qualitative understanding of the underlying physics. We show that on the background of an almost completely ionized hydrogen and helium, the decrease in  $\Gamma_1$ can be represented as a linear combination of the individual contributions of heavy elements. We also consider an inverse problem; that is, the possibility, in principle, of determining the mass fractions of the individual heavy elements from the $\Gamma_1$ profile. The inverse problem is studied in a preliminary way with a numerical experiment, because the real task needs more accurate determinations of $\Gamma_1$ from observations, which are currently not available and have to await future developments.

The plan of the paper is as follows. In Sect.~\ref{Sect_G1_Analysis} we analyze the profile of the adiabatic exponent $\Gamma_1$. The temperature range for this is selected in Sect.~\ref{Subsect_Domain}, and the sets of heavy-element mass fractions are described in Sect.~\ref{Subsect_Mixtures}. $\Gamma_1$ profiles are considered in Sect.~\ref{Subsect_G1profiles}, and the total $Z$ contribution to $\Gamma_1$ and the components of the individual contributions of the elements are defined in Sects.~\ref{Subsect_Zcontributions} and \ref{Subsect_Basis}. In Sect.~\ref{Subsect_LinearSynthesis}, we demonstrate that the total $Z$ contribution can be presented as a linear combination of a set of basis functions. In Sect.~\ref{Subsect_InverseProblem}, we consider the inverse problem, that is, an estimation of the heavy-element mass fractions based on the $\Gamma_1$ profile. The influence of the hydrogen content on the solution is studied in Sect.~\ref{Subsect_UnknownX}. In Sect.~\ref{Sect_Errors}, we consider the limited space resolution and pseudo-random errors of $\Gamma_1$ profile. The main results of the work are presented in Sect.~\ref{Sect_Conclusion}. Appendix~\ref{Appendix_Ionization} describes the physics of the decrease in $\Gamma_1$ in regions of ionization analytically.
In Appendix~\ref{Appendix_Errors}, we consider some examples of systematic and random errors.

\section{Analysis of the adiabatic-exponent profile }
\label{Sect_G1_Analysis}

\subsection{Domain of analysis}
\label{Subsect_Domain}

The $\Gamma_1$ profile is computed for the points of temperature $T$ and density $\rho $ of the solar model
 from \citep{Ayukov_Baturin_2017}, labeled 771-0001. In that paper, all model details are described. The main point is that it is a standard solar model for a high heavy-element abundance. Its depth of the convection zone is typical for high-$Z$ models ($R_{\mathrm{cz}} / R_\odot = 0.712$).
In Appendix~\ref{Appendix_Errors}, we also use the standard low-$Z$ model 771-0002, which has a shallow convection zone.

In all calculations, we use the fixed set of points $(T, \rho )$ when calculating all theoretical $\Gamma_1$ (for $Z = 0$ and for the element contributions). It is a distinctive feature of our technique.
Our method does not depend on the specific choice of a set of points $T$ and $\rho$.  A set can be chosen from a model or from proper thermodynamic calculations of the adiabat. Deviations from the adiabat are also possible.

In the framework of an ideal experiment, $\Gamma_1$ for different mixtures $Z$ was calculated on the same set of points. When a different, but fixed, profile $(T, \rho )$ is used, the result of least-square decomposition  (Sect.~\ref{Subsect_InverseProblem}) will be the same.

In the general case, however, the function studied could correspond to $\Gamma_1$ calculated at an arbitrary set of points $(T, \rho )$. Within the framework of our approach, the variations of points $(T, \rho )$ are represented by perturbations of $\Gamma_1$. We performed experiments with  $(T, \rho )$ sets that  were taken from the two standard models (Appendix~\ref{Appendix_TRho}).

We use, on the one hand, equation-of-state tables with selected mixtures of heavy elements and, on the other hand, the tables that contain only one heavy element at fixed mass fraction $Z=0.01$. The results are compared with those from tables computed for a pure hydrogen-helium mixture, that is $Z=0$.

\begin{figure}
        \centering
        \resizebox{\hsize}{!}{\includegraphics{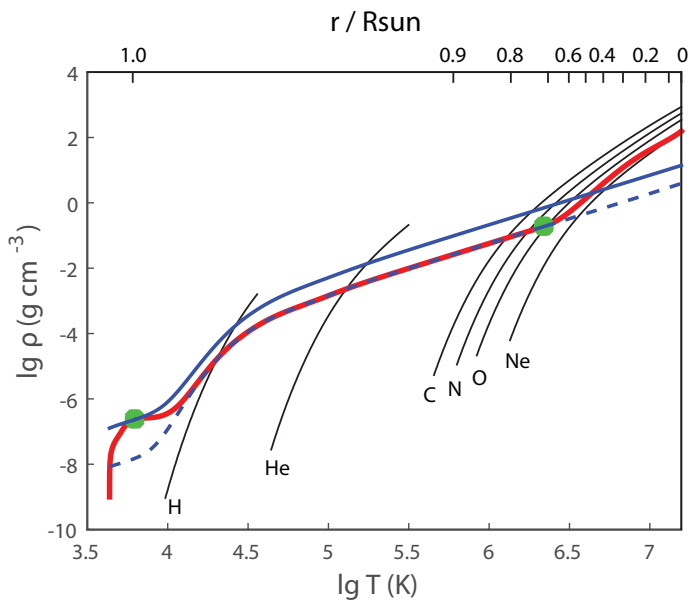}}
        \caption{Temperature and density in the standard solar model (red curve). Filled green points mark the boundaries of the convection zone. Blue curves are adiabats. Black curves show points where 50\% ionization of the last electron occurs. The top axis shows the corresponding radius in the solar model.}
        \label{FigRhoT}
   \end{figure}

The $\left( T,\rho  \right)$ points from the solar model are shown by the red line in
Fig.~\ref{FigRhoT}. Boundaries of the solar convection zone (CZ) are marked by filled green points. The convection zone becomes virtually adiabatic at $\lg T>4.3$, so entropy is constant. However, entropy varies in the colder outer layers, where convection is inefficient. To illustrate this fact, two limit adiabats are shown in Fig.~\ref{FigRhoT} by the blue curves, with all points of the convection zone lying between them.

The black curves in Fig.~\ref{FigRhoT} show the points where the 50\% ionization of the innermost electron occurs.  For hydrogen and helium, this is in the outer layers of the CZ. However, the final ionization of carbon, nitrogen, and oxygen happens in the lower part of the CZ.  Neon ionizes below the CZ.

All $\Gamma_1$ profiles in our work have been analyzed for an interval of temperatures between $\lg T=5.4$ and 6.4. Hydrogen and helium are near fully ionized under these conditions.  Furthermore, the chosen temperature interval almost lies in the adiabatic part of the CZ.

There are no exact natural limits for the boundaries of the interval. Of course, such a choice is always a compromise. That the interval contains points inside the CZ is not a strict requirement for the method, though it is desirable, as deviations from the adiabat can introduce some errors. In the used standard high-$Z$ model, the boundary of the CZ is located at $\lg T=6.344$. The temperature range has been slightly expanded to $6.4$ to include the neon ionization zone. Neon would be too important to be ignored.
        
The same arguments are applied to the low-temperature boundary. It is not possible to remove the H and He ionization zone completely. A shrinking of the interval would cause a deterioration of the basis due to an increase of linear dependence of the basic functions. We also left the extended range to include an example of neon L-shell ionization.

Undoubtedly, narrowing the interval reduces the contribution of incomplete ionization of hydrogen and helium, but our choice of boundaries is a compromise between the stability of the basis when calculating the $Z$ contribution and the linear independence of the basis. In other words, by narrowing the interval, we strengthen the linear dependence of the basis functions (see Sect.~\ref{Subsect_Zcontributions}).

\subsection{Mixtures of heavy elements}
\label{Subsect_Mixtures}
        
We analyze the $\Gamma_1$ profile for mass fractions of hydrogen $X=0.7$, helium $Y=0.28$, and heavy elements $Z=0.02$. Obviously, the abundances of the individual heavy elements satisfy the relation $\sum\nolimits_{i}{{{Z}_{i}}}=Z$. The SAHA-S equation of state takes into account all ionization stages of all eight heavy elements considered (C, N, O, Ne, Mg, Si, S, and Fe). The relative mass fractions ${{A}_{i}}={{Z}_{i}}/Z$ of the four mixtures studied are presented in Table~\ref{TabMixes}. The mass fractions ${{Z}_{i}}$ are also shown, since they are needed to analyze the results of the computations.

\begin{table*}
        \caption{Relative mass fractions ${{A}_{i}}$ and absolute mass fractions ${{Z}_{i}}=Z\cdot {{A}_{i}}$ for the heavy elements used in our mixtures with $Z=0.02$.}
\centering
\begin{tabular}{|l| *{4}{c|c|} }
        \hline \\[-3mm]
        & \multicolumn{2}{c|}{AGSS09}  & \multicolumn{2}{c|}{GN93} & \multicolumn{2}{c|}{OPAL} & \multicolumn{2}{c|}{SAS5} \\          
        \cline{2-9}
        Elem.& $A_i$ & $Z_i,\times 10^{-2}$ &  $A_i$ & $Z_i,\times 10^{-2}$ &  $A_i$ & $Z_i,\times 10^{-2}$ &   $A_i$ & $Z_i,\times 10^{-2}$ \\
        \hline
        C &     0.1816 & 0.3633 &  0.1772 &     0.3544 &  0.1907        & 0.3813 &  0.2729 &      0.5459 \\
        N &     0.0532 & 0.1064 &  0.0543 &     0.1087 &  0.0558        & 0.1117 &  0.0686 &      0.1371 \\
        O &     0.4403 & 0.8805 &  0.4932 &     0.9864 &  0.5430        & 1.0860 &  0.4967 &      0.9934 \\
        Ne&     0.0965 & 0.1930 &  0.0986 &     0.1972 &  0.2105        & 0.4210 &  0.0432 &      0.0863 \\
        Mg&     0.0544 & 0.1087 &  0.0384 &     0.0768 &        0           & 0       &  0.0404 &        0.0807 \\
        Si&     0.0511 & 0.1021 &  0.0414 &     0.0829 &        0           & 0       &  0.0328 &        0.0656 \\
        S &     0.0237 & 0.0475 &  0.0216 &     0.0432 &        0           & 0       &  0.0134 &        0.0267 \\
        Fe&     0.0992 & 0.1984 &  0.0751 &     0.1503 &        0           & 0       &  0.0321 &        0.0641 \\
        \hline
\end{tabular}
\label{TabMixes}
\end{table*}

Mixture AGSS09 is obtained by proportionally scaling the contents of the eight heavy elements from the mixture by \citet{Asplund_2009}. Mixture GN93 is obtained similarly by using the contents of \citet{Grevesse_Noels_1993}. The SAHA-S equation of state also includes  a table computed specially for a comparison with the widely known, often used, OPAL equation of state \citep{Rogers_1996, Rogers_Nayfonov_2002}. The OPAL mixture reproduces the mass fractions used in the OPAL equation of state. It contains only four heavy elements (C, N, O, and Ne). SAS5 is an artificial mixture with an increased carbon content. All mixtures differ in the relative contents of the heavy elements, whereas the total mass fraction is the same for all: $Z=0.02$ . Therefore, the mixtures differ in that the abundances of  individual heavy elements are being changed, but the total mass of all heavy elements considered remains unchanged.
Oxygen is the most abundant heavy element in the solar plasma. In AGSS09 and GN93, carbon takes the second place and neon the third. In SAS5, carbon and nitrogen are in the second and third places, respectively. In OPAL, all elements heavier than neon are treated as neon, which thus has a fictitious abundance. That way, in OPAL, neon is in the second, and carbon in the third place.

\subsection{Profile of  $\Gamma_1$ in the solar model }
\label{Subsect_G1profiles}

\begin{figure}
        \centering
        \resizebox{\hsize}{!}{\includegraphics{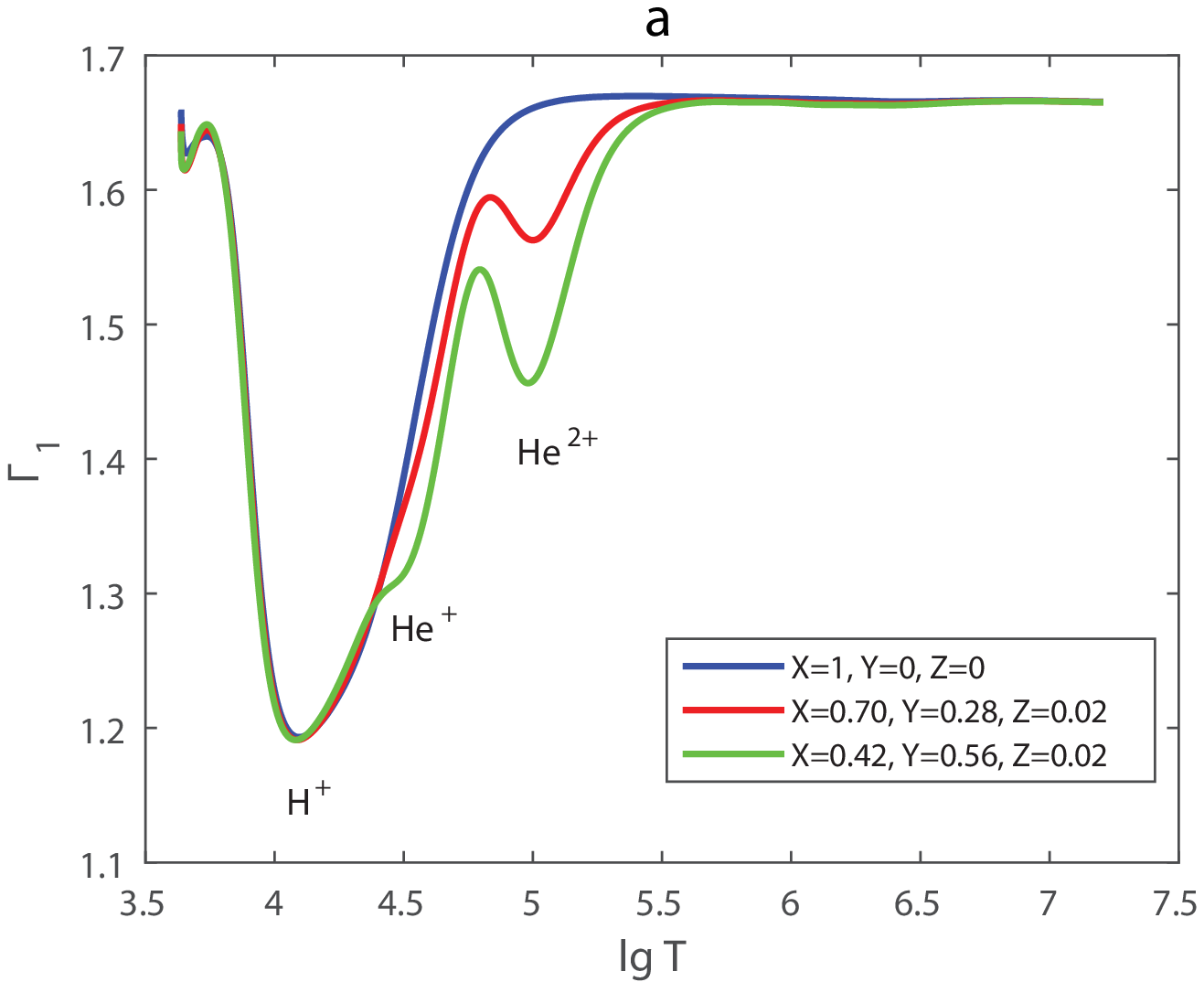}}
        \resizebox{\hsize}{!}{\includegraphics{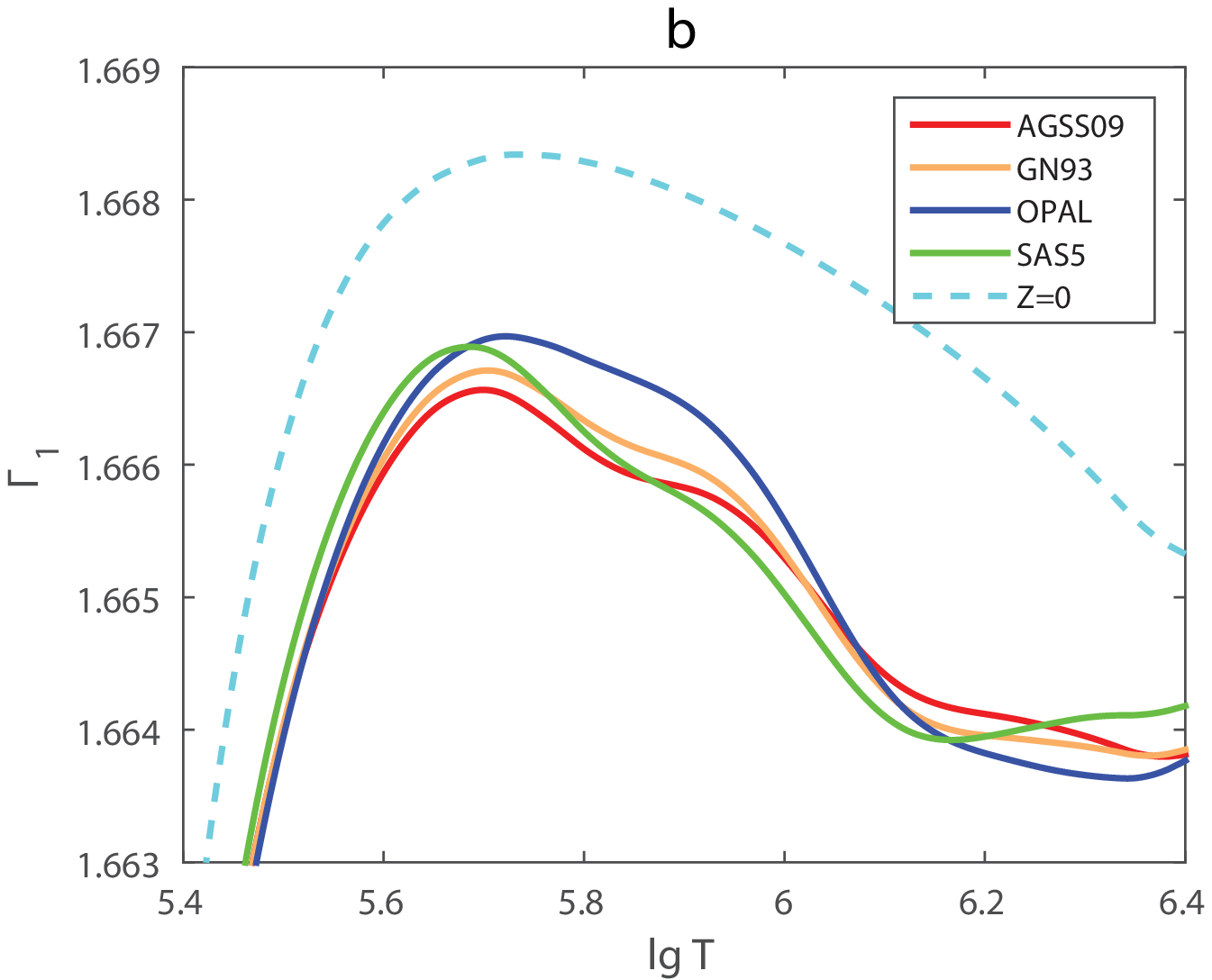}}
        \caption{Adiabatic exponent $\Gamma_1$ for points $(T,\rho )$ of the standard solar model for different chemical composition of plasma. (a) Effect of helium. (b) Effect of different mixtures of heavy elements.}
        \label{FigG1lgT}
\end{figure}

The theoretical profile of $\Gamma_1$ for the points $\left( T,\rho  \right)$ of the solar model is represented by the red curve in Fig.~\ref{FigG1lgT}a. The adiabatic exponent is equal to 5/3 in a perfect gas. It decreases in regions of ionization of atoms and ions. The maximum decrease occurs at temperatures $\lg T=4 - 5$ due to the ionization of hydrogen and helium. The blue and green curves illustrate the effect of ionization of the light elements. The blue curve is computed for pure hydrogen, and the green curve is for a plasma in which the helium content is twice that of the Sun.

We focused on fine, wave-like perturbations with an amplitude of about $3\times {{10}^{-3}}$ at temperatures between $\lg T=5.4$ and 6.4. This range is presented at a larger scale in Fig.~\ref{FigG1lgT}b. The light blue dashed curve describes $\Gamma_1$ for a hydrogen-helium plasma.
The excess of $\Gamma_1$ over $5/3$ arises due to the nonideal effect from Debye screening, which can lead to an increase in the elasticity of the plasma. The physical mechanism of this phenomena in the solar CZ is described in  \citet{Baturin_2000}.

 The two-percent addition of heavy elements to the mixture decreases $\Gamma_1$. This decrease is different for the four different mixtures (AGSS09, GN93, OPAL, and SAS5), which allows us to use the profile of $\Gamma_1$ to determine the detailed chemical composition of the plasma.

\subsection{Definition of $Z$ contribution to $\Gamma_1$ }
\label{Subsect_Zcontributions}

\begin{figure}
        \centering
        \resizebox{\hsize}{!}{\includegraphics{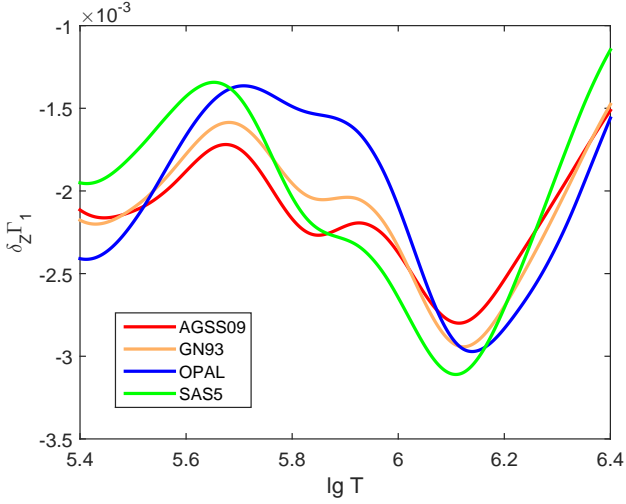}}
        \caption{$Z$ contribution ${{\delta }_{Z}}{{\Gamma }_{1}}$ of 2\% admixtures AGSS09, GN93, OPAL, and SAS5.}
        \label{Fig_dG1mixes}
\end{figure}

To understand the nature of these perturbations, let us consider the difference ${{\delta }_{Z}}{{\Gamma }_{1}}$ of $\Gamma_1$ computed for a plasma with a mixture of heavy elements ($X=0.70,Y=0.28,Z=0.02$) and $\Gamma_1$ for a pure hydrogen-helium plasma ($X=0.70,Y=0.30,Z=0$):
\begin{equation}
{{\delta }_{Z}}{{\Gamma }_{1}}={{\Gamma }_{1}}-\Gamma _{1}^{\mathrm{HHe}}.
\label{Eq_dG1}
\end{equation}
In the following, we use the term $Z$ contribution to denote this difference. It is our tool to disentangle the contribution of the heavy elements (Fig.~\ref{Fig_dG1mixes}).
The $Z$ contribution has a negative sign in the temperature range considered. The magnitude of ${{\delta }_{Z}}{{\Gamma }_{1}}$ is few fractions of a percent at $Z=0.02$. The important feature of $Z$ contribution is dependence on the chemical composition of plasma. Figure~\ref{Fig_dG1mixes} shows the $Z$ contributions for different mixtures.
Analyzing the $Z$ contribution ${{\delta }_{Z}}{{\Gamma }_{1}}$  instead of the full $\Gamma_1$ plays a fundamental role. Nonideal effects, interpolation errors, and some effects of hydrogen and helium ionization are subtracted in this case.  The $Z$ contribution is related to the ionization of a small admixture on the background of the constant ionization of hydrogen and helium. By ``small admixture'' we mean that the number of electrons added by the element under consideration is small compared to all electrons in the plasma. Thus, we study their ionization over the background of almost fully ionized hydrogen and helium, which provides the bulk of the electrons.

The effect of the equation of state generally consists of two parts. One is caused by ideal ionization, the other by nonideal effects, such as the Coulomb correction and others. By maintaining the equation of state constant, the $Z$ contributions allow us to subtract most of the Coulomb correction. Thus, we amplify the effect of the heavy-element ionization. If in the same study one had also used another equation of state for the $\Gamma_1$ contribution, the nonideal part could not have been subtracted. However, since we succeed in doing that, in this sense we can claim independence of the equation of state.

In the current version of the method, a theoretical background profile of $\Gamma_1^{\mathrm{HHe}}$       is computed for known $X$,      fixed points $(T, \rho )$, in frames of one equation of state. Certainly, if the nature of $\Gamma_1$ is unknown, the true background profile differs from the suggested theoretical profile, and then additional errors will enter our method. These errors are an unavoidable price for the increasing sensitivity to contributions of heavy elements. We give some examples of a possible variation of $\Gamma_1^\mathrm{HHe}$ in Appendix~\ref{Appendix_G1_HHe}. We provide an example for an alternative equation of state (OPAL) as well (see Appendix~\ref{Appendix_OPALeos}).

\subsection{Basis of the $Z$ contributions of elements}
\label{Subsect_Basis}

\begin{figure}
        \centering
        \resizebox{\hsize}{!}{\includegraphics{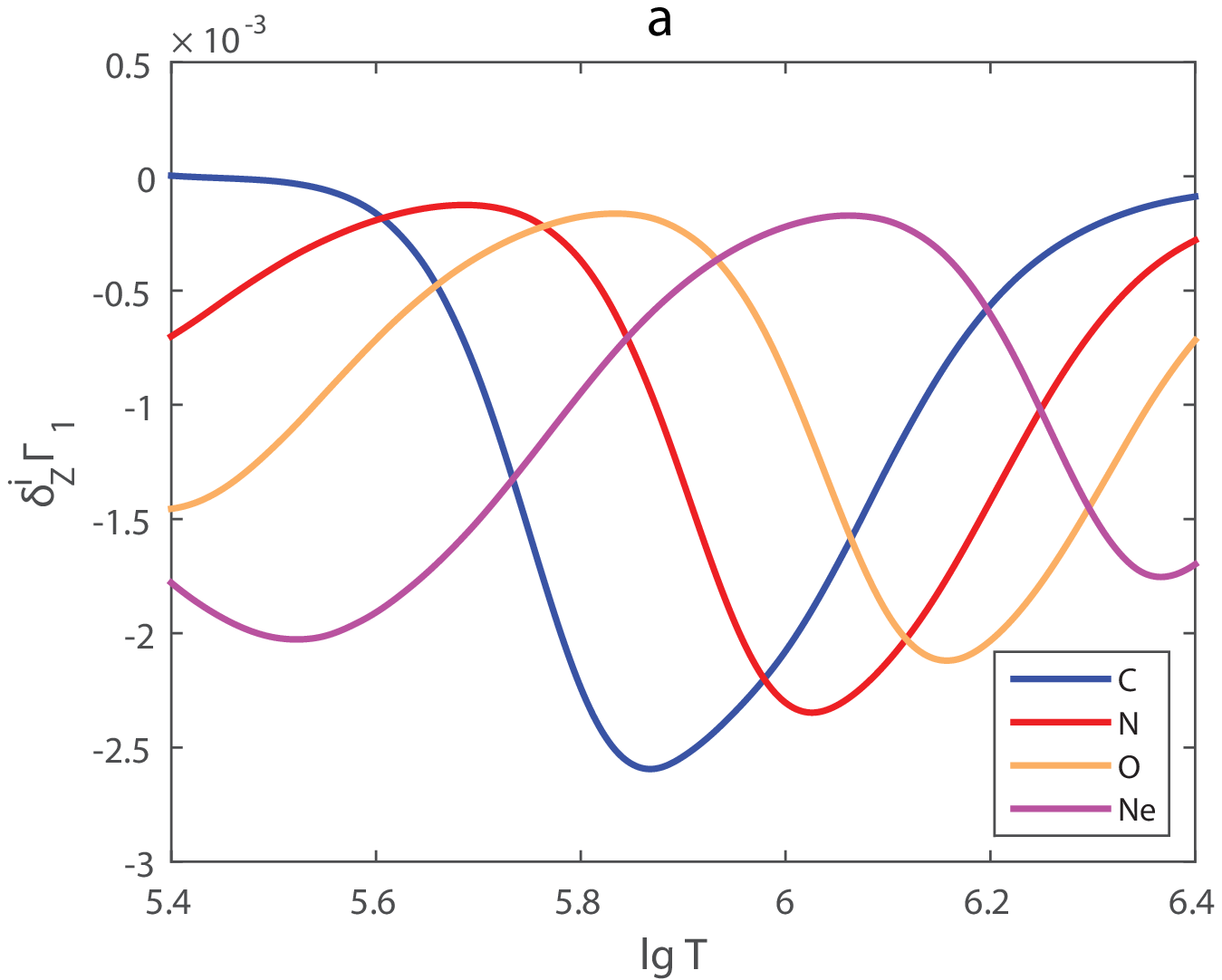}}
        \resizebox{\hsize}{!}{\includegraphics{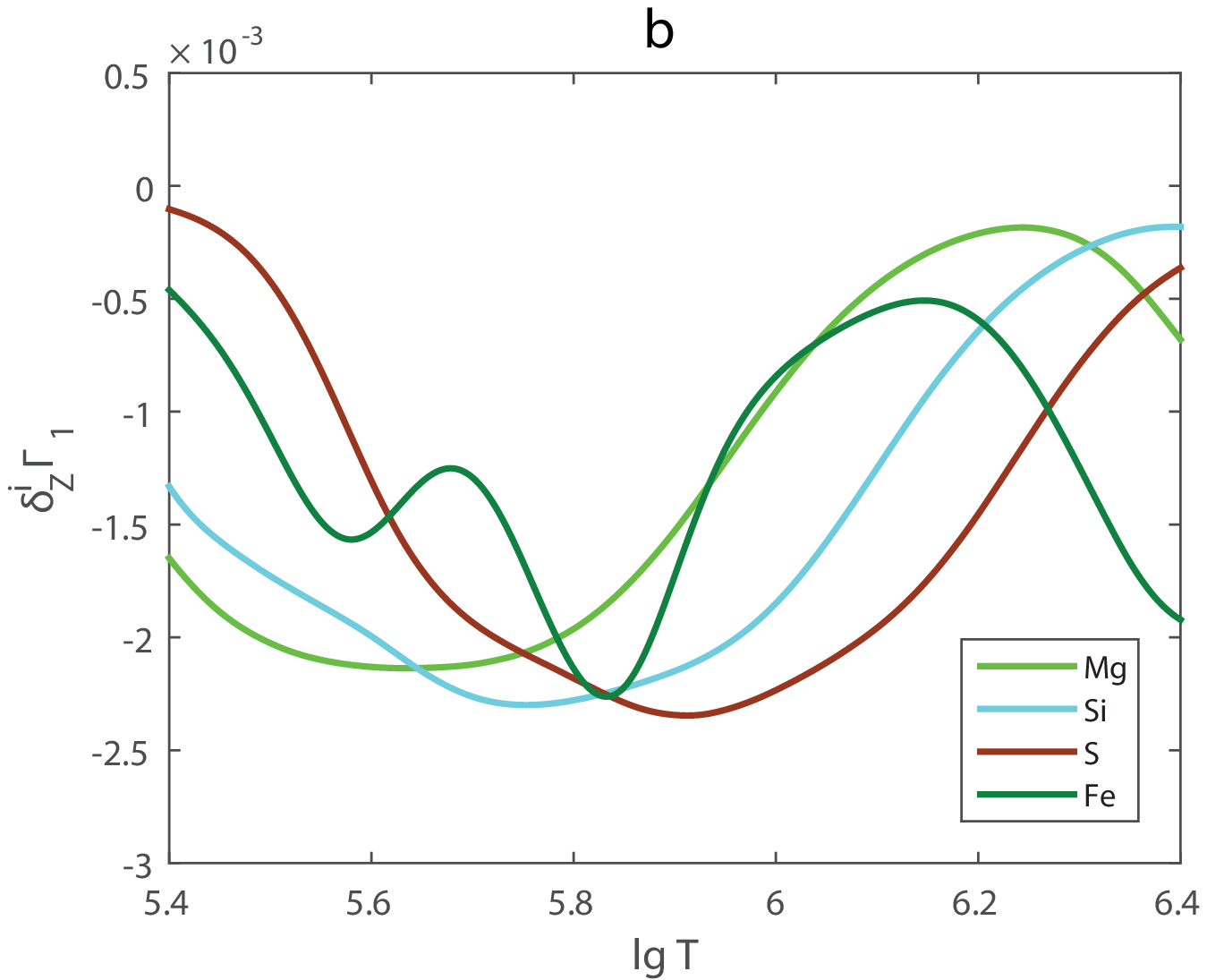}}
        \caption{Basis {$\delta _{Z}^{i}{{\Gamma }_{1}}$ } of individual contributions to ${{\delta }_{Z}}{{\Gamma }_{1}}$.}
        \label{Fig_dG1basis}
\end{figure}

We now consider individual $Z$ contributions $\delta _{Z}^{i}{{\Gamma }_{1}}(X,Y,{{Z}_{i}}^{\mathrm{b}})$ of eight heavy elements available in the SAHA-S equation of state (Fig.~\ref{Fig_dG1basis}). Here, the index ``b'' in variable $Z_{i}^{\mathrm{b}}$  means ``basis'', and $i=1...8$ labels a heavy element (C, N, O, etc.). The contributions are obtained using special computations for plasma, in which one percent of mass is represented by only one heavy element (${{Z}_{i}}^{\mathrm{b}}=0.01$), and the rest of the masses are hydrogen $(X=0.70)$  and helium ($Y=0.29$). So, we obtain adiabatic exponent $\Gamma _{1}^{i}$ and then subtract a background that is $\Gamma _{1}^{\rm HHe}$ for  hydrogen-helium plasma ($X=0.70,Y=0.30,Z=0$):
\begin{equation}
\delta _{Z}^{i}{{\Gamma }_{1}}=\Gamma _{1}^{i}-{{\Gamma }_{1}}^{\rm HHe}.
\label{Eq_Basis}                
\end{equation}

The calculation of the $Z$ contribution can be performed either for a fixed $X$ or for a fixed $Y$. If $Y$ is fixed, the error of the linear synthesis (see the following section) increases by about two times in comparison with fixed $X$. It may be explained by the simultaneous incomplete ionization of hydrogen and helium at the low-temperature edge of the interval. Since the number of helium particles is about ten times less than the number of hydrogen particles, the effect of the incomplete hydrogen ionization is stronger.
The $\Gamma_1$ profile is explained qualitatively by ionizations (see Appendix~\ref{Appendix_Ionization}).

For the most abundant elements (C, N, O, and Ne) in the adiabatic part of the convection zone, the main role is played by ionization corresponding to detachment of the last two electrons from the inner, so-called K, shell. For neon and heavier elements, a contribution of L-shell ionization falls into the range being investigated.

Every curve has a clear minimum at a certain temperature. That is why these functions are convenient to use as a basis of decomposition.

\subsection{Linear synthesis of the $Z$ contribution}
\label{Subsect_LinearSynthesis}

\begin{figure}
        \centering
        \resizebox{\hsize}{!}{\includegraphics{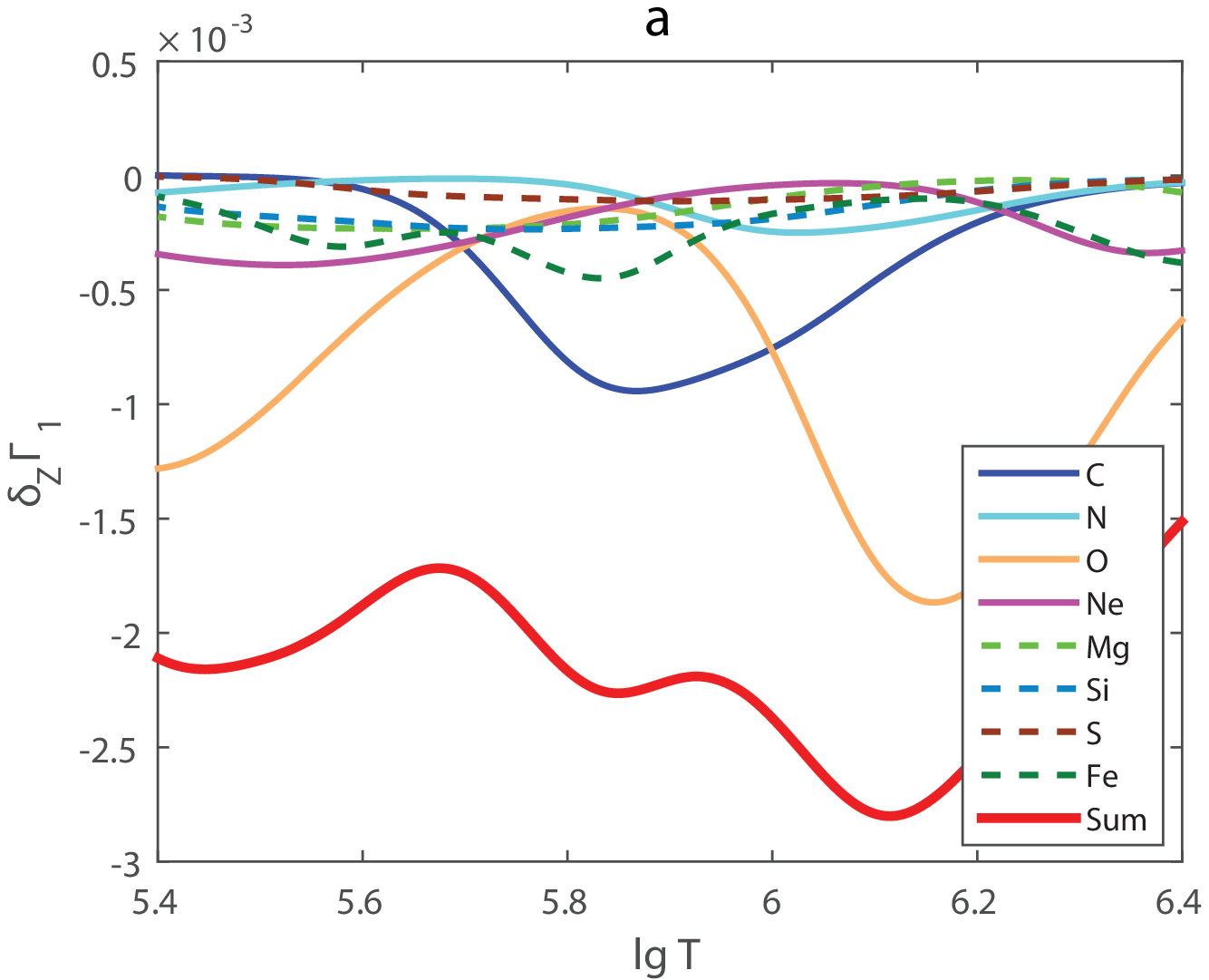}}
        \resizebox{\hsize}{!}{\includegraphics{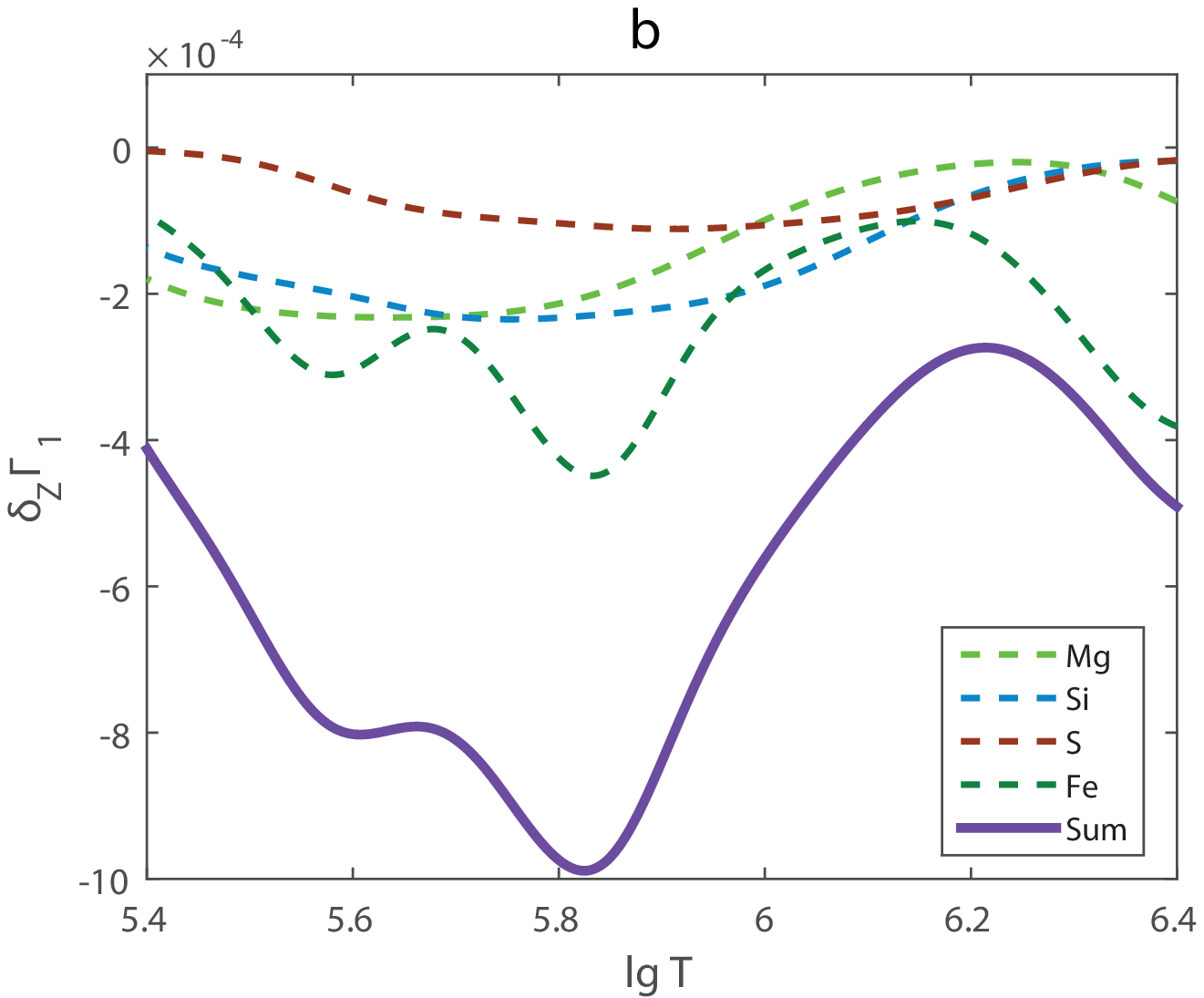}}
        \caption{Weighted contributions ${{Z}_{i}}\cdot \delta _{Z}^{i}{{\Gamma }_{1}}$  of individual elements and their sum ${{\delta }_{Z}}{{\Gamma }_{1}}^{\mathrm{LC}}$ for the  AGSS09 mixture. (a) Contributions of all eight considered elements. (b) Contributions of Mg, Si, S, and Fe at a larger scale.}
        \label{Fig_dG1_elementsVzvesh}
\end{figure}

Straight modeling of the $Z$ contribution ${{\delta }_{Z}}{{\Gamma }_{1}}$ consists of the computation of the following linear combination (LC hereafter):
\begin{equation}
\delta _{Z}^{{}}{{\Gamma }_{1}}^{\mathrm{LC}}=\sum\limits_{i}{\left( {{{Z}_{i}}}/{\Delta {{Z}^{\mathrm{b}}}}\; \right)\cdot \text{ }}\delta _{Z}^{i}{{\Gamma }_{1}}
\label{Eq_LC}
,\end{equation}
using known mass fractions ${{Z}_{i}}$, basis functions $\delta _{Z}^{i}{{\Gamma }_{1}}$ (Eq.~(\ref{Eq_Basis}), and $\Delta {{Z}^{\mathrm{b}}}=0.01$. Fig.~\ref{Fig_dG1_elementsVzvesh}a shows all eight terms of the sum~(\ref{Eq_LC}) and the linear combination ${{\delta }_{Z}}{{\Gamma }_{1}}^{\mathrm{LC}}$(thick red line) for the mixture AGSS09. The most abundant elements (oxygen, carbon, neon, and nitrogen) give the main contributions.

An analysis of the contributions in Fig.~\ref{Fig_dG1_elementsVzvesh}a allows us to interpret the shape of the resultant profile.
Oxygen and neon create the local minimum at $\lg T=5.4$. Carbon produces the minimum at $\lg T=5.84,$ and oxygen  causes the decrease at $\lg T=6.1$.

Figure~\ref{Fig_dG1_elementsVzvesh}b shows contributions of magnesium, silicon, sulfur, and iron at the larger scale. The violet curve is their sum. The total contribution of these elements reaches ${{10}^{-3 }}$ , which is comparable to the contribution of carbon. The contribution of iron is dominant in the group. Its minimum almost coincides with carbon, which complicates the problem of differential analysis. The OPAL equation of state does not include these elements.

\begin{figure}
        \centering
        \resizebox{\hsize}{!}{\includegraphics{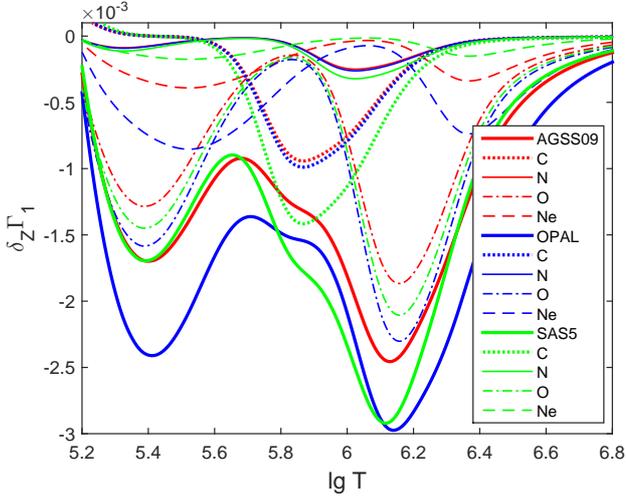}}
        \caption{Weighted contributions ${{Z}_{i}}\cdot \delta _{Z}^{i}{{\Gamma }_{1}}$ of main elements (C, N, O, and Ne) in ${{\delta }_{Z}}{{\Gamma }_{1}}$ for mixtures AGSS09, OPAL, and SAS5.}
        \label{Fig_dG1_elementsVzvesh_mixes}
\end{figure}

Figure~\ref{Fig_dG1_elementsVzvesh_mixes} illustrates the linear synthesis of $\delta _Z\Gamma _1$
for three mixtures: AGSS09, OPAL, SAS5. It shows only contributions of the most abundant elements (C, N, O, and Ne) and their sum without the contribution of the heavier group. The figure
demonstrates that variations of contents in the C-Ne group can significantly change the total Z-profile. The figure allows us also to compare the content of elements in the considered mixtures. For example, enhanced content of neon in OPAL is the main feature of the mixture. The direct effect of neon can be
noticed at $\lg T=5.4$  and at $\lg T=6.4$; oxygen contribution is also significant there. The content of oxygen and neon is more prominent in OPAL than in other mixtures, which results in deeper total $Z$ contribution (blue solid thick curve). Similarly, a large percentage of carbon in mixture SAS5 produces a relative decrease at $\lg T=5.84$ (green solid thick curve), which is minor in other mixtures. However, this carbon minimum is influenced by the heavy group, first of all by iron, and can disappear in the total curve.

\begin{figure}
        \centering
        \resizebox{\hsize}{!}{\includegraphics{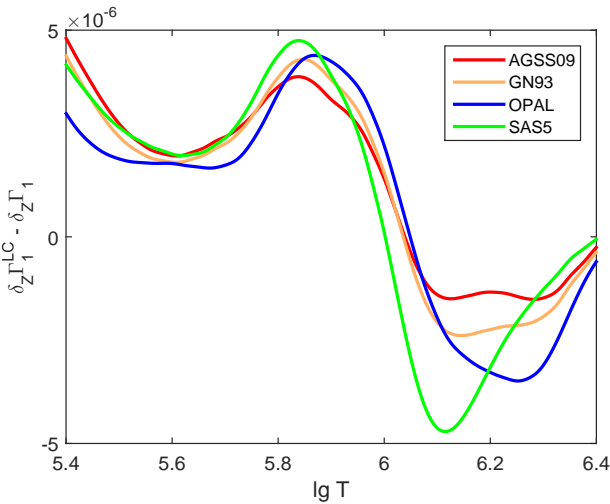}}
        \caption{Difference between the linear combination ${{\delta }_{Z}}{{\Gamma }_{1}}^{\mathrm{LC}}$ and exact $Z$ contribution ${{\delta }_{Z}}{{\Gamma }_{1}}$ for  mixtures AGSS09, GN93, OPAL, and SAS5.}
        \label{Fig_DdG1_lin_mix}
\end{figure}

We now compare the linear combination ${{\delta }_{Z}}{{\Gamma }_{1}}^{\rm LC}$ and the exact $Z$ contribution ${{\delta }_{Z}}{{\Gamma }_{1}}$ to estimate the accuracy of the linear combination. The difference is shown in Fig.~\ref{Fig_DdG1_lin_mix}. It does not exceed $5\times {{10}^{-6}}$ for all considered mixtures. The relative error is about 0.1\%, which demonstrates the reliability of the linear synthesis.  The errors of the linear synthesis for the mixtures are generally similar, and the divergence between them does not exceed $(2-3) \times {{10}^{-6}}$.

The error profiles have a peculiar character. The error is positive at low temperatures and negative at high temperatures of the range in Fig.~\ref{Fig_DdG1_lin_mix}. The physical nature of the error of the linear synthesis is related to ionization of elements in a mixture. We assume that the electron concentration in a mixture slightly differs from that upon computation of individual basis contributions. This leads to somewhat different ionization conditions for key components.
The error of the linear synthesis plays an important role in the inverse problem discussed in the next section.

\subsection{Inverse problem of least-squares decomposition}
\label{Subsect_InverseProblem}

The inverse problem consists of determining mass fractions  of heavy elements ${{Z}_{i}}$ using a given ${{\delta }_{Z}}{{\Gamma }_{1}}$ profile. We solve the problem using the method of linear, unconstrained least squares (LS).

We analyze ${{\delta }_{Z}}{{\Gamma }_{1}}$ computed in the SAHA-S equation of state for the chemical composition $X=0.70,$ $Y=0.28,$  $Z=0.02$ and different mixtures. Then, we estimate mass fractions ${{Z}_{i}}^{\mathrm{LS}}$ using the LS method. The results are presented in Table~\ref{Table_LS_Decomposition}.

\begin{table*}
        \caption{Mass fractions of heavy elements ${{Z}_{i}}^{\mathrm{LS}}$ determined from  a given $Z$ contribution ${{\delta }_{Z}}{{\Gamma }_{1}}$ , absolute error $\Delta {{Z}_{i}}$ and relative error $\Delta {{Z}_{i}}/{{Z}_{i}}$ .}
        \centering
        \begin{tabular}{|l| *{4}{c|c|c|} }
                \hline
                & \multicolumn{3}{c|}{AGSS09}  & \multicolumn{3}{c|}{GN93} & \multicolumn{3}{c|}{OPAL} & \multicolumn{3}{c|}{SAS5} \\              
                \cline{2-13}
                Elem.& $Z_i^{\mathrm{LS}},$ & $\Delta Z_i,$ & ${\Delta Z_i}/{Z_i},$ &   $Z_i^{\mathrm{LS}},$ & $\Delta Z_i,$ &  ${\Delta Z_i}/{Z_i},$ &  $Z_i^{\mathrm{LS}},$ & $\Delta Z_i,$ & ${\Delta Z_i}/{Z_i},$  &   $Z_i^{\mathrm{LS}},$ & $\Delta Z_i,$ & ${\Delta Z_i}/{Z_i},$  \\
                & $\times 10^{-2}$       & $\times 10^{-6}$ & \% & $\times 10^{-2}$   &  $ \times 10^{-6}$ & \% & $\times 10^{-2}$  &$ \times 10^{-6}$ & \% & $\times 10^{-2}$  &$\times 10^{-6}$ & \%  \\
                \hline
                C & 0.3666 & 33 & 0.9 &  0.3578 & 34 & 1.0 &  0.3830  & 17 & 0.4 &  0.5505 & 48 & 0.9 \\
                N & 0.1066 & 2 & 0.2  &  0.1092 & 5  & 0.4 &  0.1136  & 19 & 1.7 &  0.1363 & -8 & -0.6\\
                O & 0.8803 &-2 &-0.02 &  0.9856 & -8 & -0.1&  1.0841  & -19& -0.2&  0.9923 & -11& -0.1\\
                Ne& 0.1937 &  7 & 0.3 &  0.1980 & 8  & 0.4 &  0.4204  & -6 & -0.1&  0.0889 & 26 & 3.0 \\
                Mg& 0.1154 & 66 & 6.1 &  0.0830 & 61 & 8.0 &  0.0057  & 57 & -   &  0.0854 & 47 & 5.8 \\
                Si& 0.0970 &-51 & -5.0&  0.0783 & -45& -5.5&  -0.0032 & -32& -   &  0.0616 & -40& -6.1\\
                S & 0.0472 & -3 & -0.6&  0.0425 & -7 & -1.6&  -0.0012 & -12& -   &  0.0256 & -11& -4.2\\
                Fe& 0.1961 & -24& -1.2&  0.1481 & -21& -1.4&  -0.0006 & -6 & -   &  0.0614 & -27& -4.2\\
                \hline
                $Z^{\mathrm{LS}}$& \multicolumn{3}{c|}{0.020028}  & \multicolumn{3}{c|}{0.020026} & \multicolumn{3}{c|}{0.020018} & \multicolumn{3}{c|}{0.020024} \\
                \hline
                $\eta$& \multicolumn{3}{c|}{$2.320\times 10^{-6}$}  & \multicolumn{3}{c|}{$2.525\times 10^{-6}$} & \multicolumn{3}{c|}{$2.704\times 10^{-6}$} & \multicolumn{3}{c|}{$2.996\times 10^{-6}$} \\
                \hline
                $\varepsilon$& \multicolumn{3}{c|}{$2.513\times 10^{-7}$} & \multicolumn{3}{c|}{$2.778\times 10^{-7}$} & \multicolumn{3}{c|}{$2.280\times 10^{-7}$}& \multicolumn{3}{c|}{$5.249\times 10^{-7}$}\\
                \hline
                $\sigma$ & \multicolumn{3}{c|}{$2.307\times 10^{-6}$} & \multicolumn{3}{c|}{$2.509\times 10^{-6}$} & \multicolumn{3}{c|}{$2.694\times 10^{-6}$}& \multicolumn{3}{c|}{$2.949\times 10^{-6}$}\\
                \hline
        \end{tabular}
        \tablefoot{The results are obtained from the decomposition using all eight considered elements, whereas the OPAL mixture includes only four elements.}
        \label{Table_LS_Decomposition}
\end{table*}

Table~\ref{Table_LS_Decomposition} shows absolute and relative differences between the computed values ${{Z}_{i}}^{\mathrm{LS}}$ and original mass fractions ${{Z}_{i}}$ (see Table~\ref{TabMixes}):

\begin{equation}
\Delta {{Z}_{i}}={{Z}_{i}}^{\mathrm{LS}}-{{Z}_{i}}, \qquad
\frac{\Delta {{Z}_{i}}}{{{Z}_{i}}}=\frac{{{Z}_{i}}^{\mathrm{LS}}-{{Z}_{i}}^{{}}}{{{Z}_{i}}^{{}}}\cdot 100\%.
\label{Eq_dZ}
\end{equation}

The results demonstrate good agreement between the values. For the AGSS09  mixture, discrepancies are smaller or about one percent. The exceptions are Mg and Si, for which relative differences reach 5-6\%.   The errors have opposite signs and compensate each other in the total sum. The LS method estimates the mass fraction of all heavy elements,  ${{Z}^{\mathrm{LS}}}=0.020028,$ which is close to the original value $Z=0.02$. In the case of GN93, the errors are similar. The OPAL mixture includes only four elements: carbon, nitrogen, oxygen, and neon; the accuracy is several tenths of a percent for C, O, and Ne, and 1.7\% for N. The negative coefficients of Si, S, and Fe have no physical meaning and represent an error in the method. In the SAS5 mixture, the errors for carbon, nitrogen, and oxygen are fractions of a percent. For the rest elements, they reach several percent. The total mass fraction of heavy elements $Z$ is estimated with an accuracy of one tenth of a percent in all considered cases. Thus, we conclude that the LS method gives an accurate estimation for mass fractions of heavy elements under the ideal condition that we know the percentages of hydrogen and helium, which are needed to compute the background $\Gamma _{1}^{\mathrm{HHe}}$  and basis $\left\{ \delta _{Z}^{i}{{\Gamma }_{1}} \right\}$ .

\begin{figure}
        \centering
        \resizebox{\hsize}{!}{\includegraphics{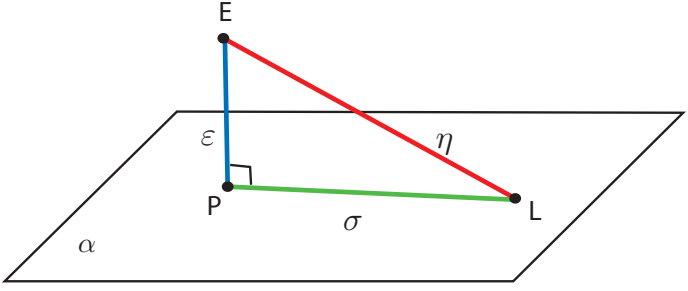}}
        \caption{Geometrical interpretation of the least-squares method.}
        \label{Fig_LS_shema}
\end{figure}

Figure~\ref{Fig_LS_shema} schematically presents a geometrical interpretation of the LS method. Every point in the figure corresponds to a certain ${{\delta }_{Z}}{{\Gamma }_{1}}\left( T \right)$.  Point $E$ corresponds to the exact $Z$ contribution ${{\delta }_{Z}}{{\Gamma }_{1}}^{{}}$, computed in the equation of state for a mixture of elements. Plane $\alpha $ is a subspace of all linear combinations and its dimension equals  the number of basis elements. Every point from subspace $\alpha $ corresponds to vector of mass fractions $\left\{ {{{\tilde{Z}}}_{i}} \right\}$.

Point $E$ does not belong to subspace $\alpha $ in general; that is, no linear combination can fully accurately reproduce the $Z$ contribution ${{\delta }_{Z}}{{\Gamma }_{1}}^{{}}$. There are two particular points in subspace $\alpha $. Point $L$ corresponds to linear combination ${{\delta }_{z}}{{\Gamma }_{1}}^{\mathrm{LC}}$ with original values ${{Z}_{i}}^{{}}$ from Table~\ref{TabMixes}. Point $P$ describes linear combination ${{\delta }_{z}}{{\Gamma }_{1}}^{\mathrm{LS}}$  with values $\left\{ {{Z}_{i}}^{\mathrm{LS}} \right\}$ obtained using the LS method. Point $P$ is the nearest point in subspace $\alpha $ to $E$ by definition of the LS method. Segment $EP$ is orthogonal to $\alpha $, because the LS method finds orthogonal  projection of point $E$ onto subspace $\alpha $.
Every segment is a difference of $Z$ contributions. The segments can be characterized by Euclidean norms (distance) computed as standard deviations of $Z$ profiles. For the three segments in Fig.~\ref{Fig_LS_shema}, they are

\begin{equation}
\eta ={{\left\| EL \right\|}_{2}}=\sqrt{\frac{1}{N}\sum\limits_{j=1}^{N}{{{[{{\delta }_{z}}{{\Gamma }_{1}}^{\mathrm{LC}}-{{\delta }_{Z}}{{\Gamma }_{1}}]}^{2}}}},
\label{Eq_eta}
\end{equation}  
\begin{equation}
\varepsilon ={{\left\| EP \right\|}_{2}}=\sqrt{\frac{1}{N}\sum\limits_{j=1}^{N}{{{[{{\delta }_{z}}{{\Gamma }_{1}}^{\mathrm{LS}}-{{\delta }_{Z}}{{\Gamma }_{1}}]}^{2}}}},
\label{Eq_epsilon}
\end{equation}
\begin{equation}
\sigma ={{\left\| PL \right\|}_{2}}=\sqrt{\frac{1}{N}\sum\limits_{j=1}^{N}{{{[{{\delta }_{z}}{{\Gamma }_{1}}^{\mathrm{LC}}-{{\delta }_{Z}}{{\Gamma }_{1}}^{\mathrm{LS}}]}^{2}}}}.
\label{Eq_sigma}
\end{equation}

\noindent
Here, summation is performed over all points $({{T}_{j}},{{\rho }_{j}}),\text{ }j=1...N$ of the solar model (Fig.~\ref{FigRhoT}) within the $\lg T=5.4-6.4 $ range. The orthogonality of the segments results in the relation ${{\eta }^{2}}={{\varepsilon }^{2}}+{{\sigma }^{2}}.$

Segment $EL$ of length $\eta $ estimates the inaccuracy of linear synthesis that is caused by physical nonlinearity of ionization in the mixture that is considered in Sect.~\ref{Subsect_LinearSynthesis}. $\eta \approx (2-3)\times {{10}^{-6}}$(see Table~\ref{Table_LS_Decomposition}).
Segment $\varepsilon =EP$ is a distance from point $E$ to plane $\alpha $. It describes a part of $Z$ contribution that cannot be represented as a linear combination of the element contributions. Value $\varepsilon $ is mainly determined by the completeness of the basis and thus depends on subspace $\alpha $. Including additional elements expands $\alpha $ and results in a decrease in $\varepsilon $. However, if a considered temperature range is narrow, then $\alpha $ cannot be expanded because new components are linearly dependent on those already included in $\alpha $.

It is interesting to consider LS decomposition in the case of applied constraints to the space $\alpha $. Constraints may be applied by a limited number of basic elements or a requirement of nonnegative abundances, or by the known total Z content. It is important to note that the squeezing of space may lead only to an increase of least squares   $\varepsilon $. These approaches do not increase the accuracy of the LS solution and could simply help to reject physically inappropriate solutions.

Value $\sigma ={{\left\| PL \right\|}_{2}}$ characterizes an error of estimation $Z_{i}^{\mathrm{LS}}$  compared to the original values ${{Z}_{i}}$. If $\varepsilon \ll \sigma, $ then $\sigma \to \eta $. Thus, an estimation error cannot be eliminated, because a possible decrease in $\varepsilon $ does not decrease $\sigma$.

In the inverse problem, the  LS method estimates $\left\{ {{Z}_{i}}^{\mathrm{LS}} \right\}$ and  hence $\varepsilon $ (not $\eta $ or $\sigma, $ because we do not know, a priori, the original $\left\{ {{Z}_{i}} \right\}$). So, $\varepsilon $ deserves special attention.
In our test experiments, value $\varepsilon $ is six-to-ten times smaller than $\eta $. Therefore, we conclude that $\varepsilon $ is close to its minimal value and eight elements are sufficient for space $\alpha $. The obtained mass fractions and their errors are not likely to be improved by including additional elements. If  a profile  contains an additional undecomposed component, the residuals increase and the LS estimation of $\left\{ {{Z}_{i}}^{\mathrm{LS}} \right\}$ becomes unreliable (Sect.~\ref{Subsect_UnknownX}).

\subsection{Case of unknown hydrogen content}
\label{Subsect_UnknownX}

The computation of $Z$ contributions (Eqs.~(\ref{Eq_dG1}) and (\ref{Eq_Basis})) uses background profile ${{\Gamma }_{1}}^{\mathrm{HHe}}$ for hydrogen-helium plasma and profiles $\Gamma _{1}^{i}$ for different elements.  However, hydrogen content can be unknown in the investigated profile $\Gamma_1$ and hence in the theoretical functions ${{\Gamma }_{1}}^{\mathrm{HHe}}$ and in $\Gamma _{1}^{i}$. In this section, we estimate the sensitivity of the method to the hydrogen content in the studied profile ${{\delta }_{Z}}{{\Gamma }_{1}}$ and in the profiles $\delta _{Z}^{i}{{\Gamma }_{1}}$  used for decomposition.

We use the mixture AGSS09 as an example. The test function $\Gamma_1$ is computed for $X=0.70,Y=0.28,Z=0.02$, but we take $X=0.69$ for the computation of the background ${{\Gamma }_{1}}^{\mathrm{HHe}}$and basis $\left\{ \delta _{Z}^{i}{{\Gamma }_{1}} \right\}$. Thus, the content of hydrogen in the studied function differs from the background and the basis. The results of LS decomposition are presented in Table~\ref{Table_UnknownX}.

\begin{table}
        \caption{Mass fractions of heavy elements ${{Z}_{i}}^{\mathrm{LS}}$determined from the test profile $\Gamma_1$ at $X=0.70$ that differs from $X=0.69$ in the background ${{\Gamma }_{1}}^{\mathrm{HHe}}$and basis $\left\{ \delta _{Z}^{i}{{\Gamma }_{1}} \right\}$. }
        \centering
        \begin{tabular}{|l|c|c|}
                \hline
                Elem. & $Z_i^{\mathrm{LS}}, \times 10^{-2} $    & $\Delta Z_i/Z_i$, \% \\
                \hline  
                C &     0.2620  & -28 \\
                N &     0.0900  & -15 \\
                O &     0.8342  &  -5 \\
                Ne&     0.1435  & -26 \\
                Mg&     -0.0777 &-171 \\
                Si&     0.2505  & 145 \\
                S &     0.0884  &  86 \\
                Fe&     0.2931  &  48 \\
                \hline
                $Z^{\mathrm{LS}}$ & \multicolumn{2}{|c|}{0.018840} \\
                \hline  
                $\eta$   & \multicolumn{2}{|c|}{$6.925\times 10^{-5}$}   \\
                \hline  
                $\varepsilon$ & \multicolumn{2}{|c|}{$1.044\times 10^{-5}$}   \\
                \hline  
                $\sigma$ & \multicolumn{2}{|c|}{$6.846\times 10^{-5}$}   \\
                \hline
        \end{tabular}
\label{Table_UnknownX}
\end{table}

Errors  $\Delta {{Z}_{i}}/{{Z}_{i}}$ are significantly increased (see Table~\ref{Table_LS_Decomposition}). They reach tens of percent in most cases and exceed 100\% for magnesium and silicon. Moreover, a negative mass fraction $Z_i$ is obtained for magnesium that has no physical meaning. The standard deviations $\eta$, $\varepsilon$, and $\sigma$ are also increased significantly. The results show that hydrogen content in the background and basis plays a critical role.

\begin{figure}
        \centering
        \resizebox{\hsize}{!}{\includegraphics{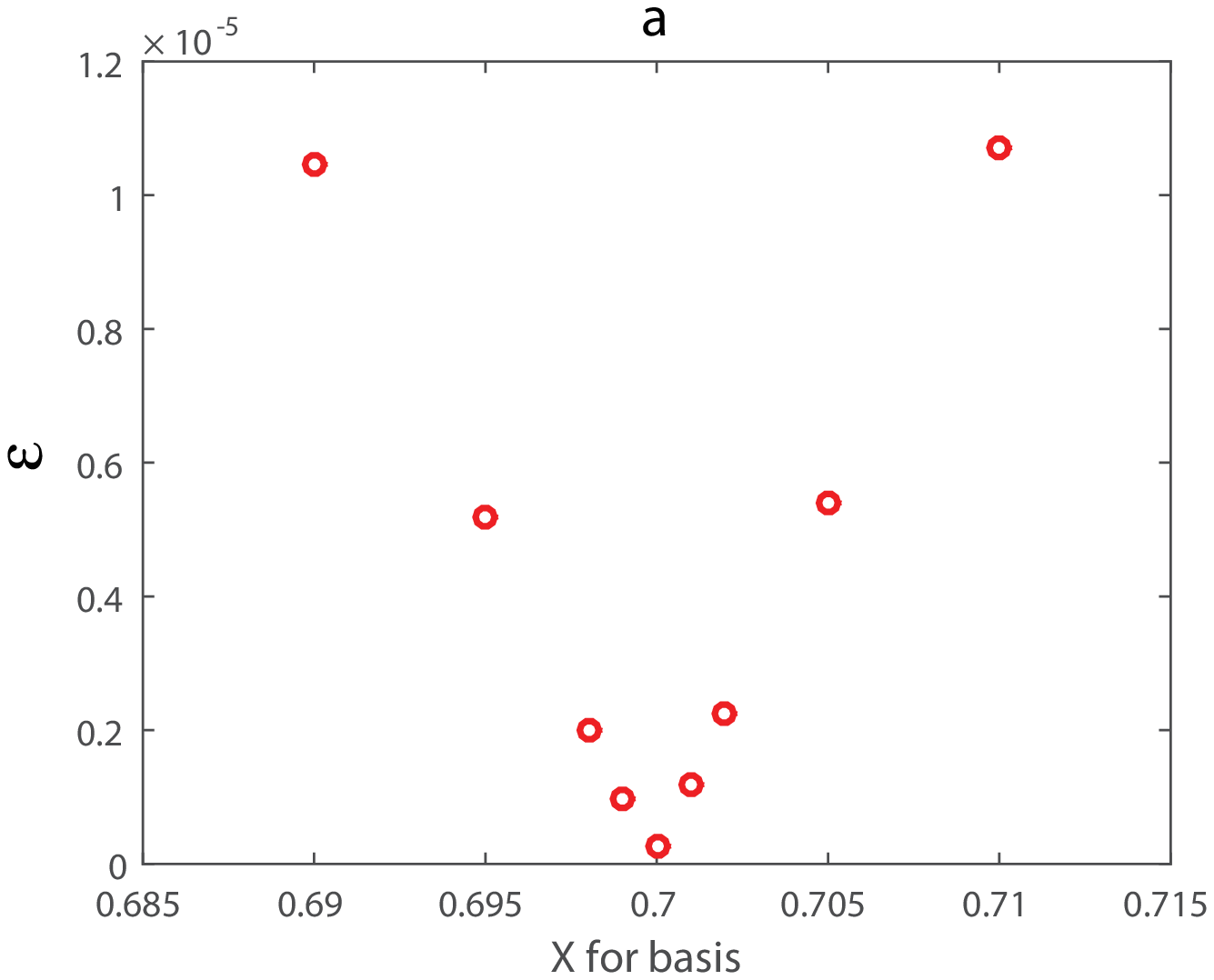}}
        \resizebox{\hsize}{!}{\includegraphics{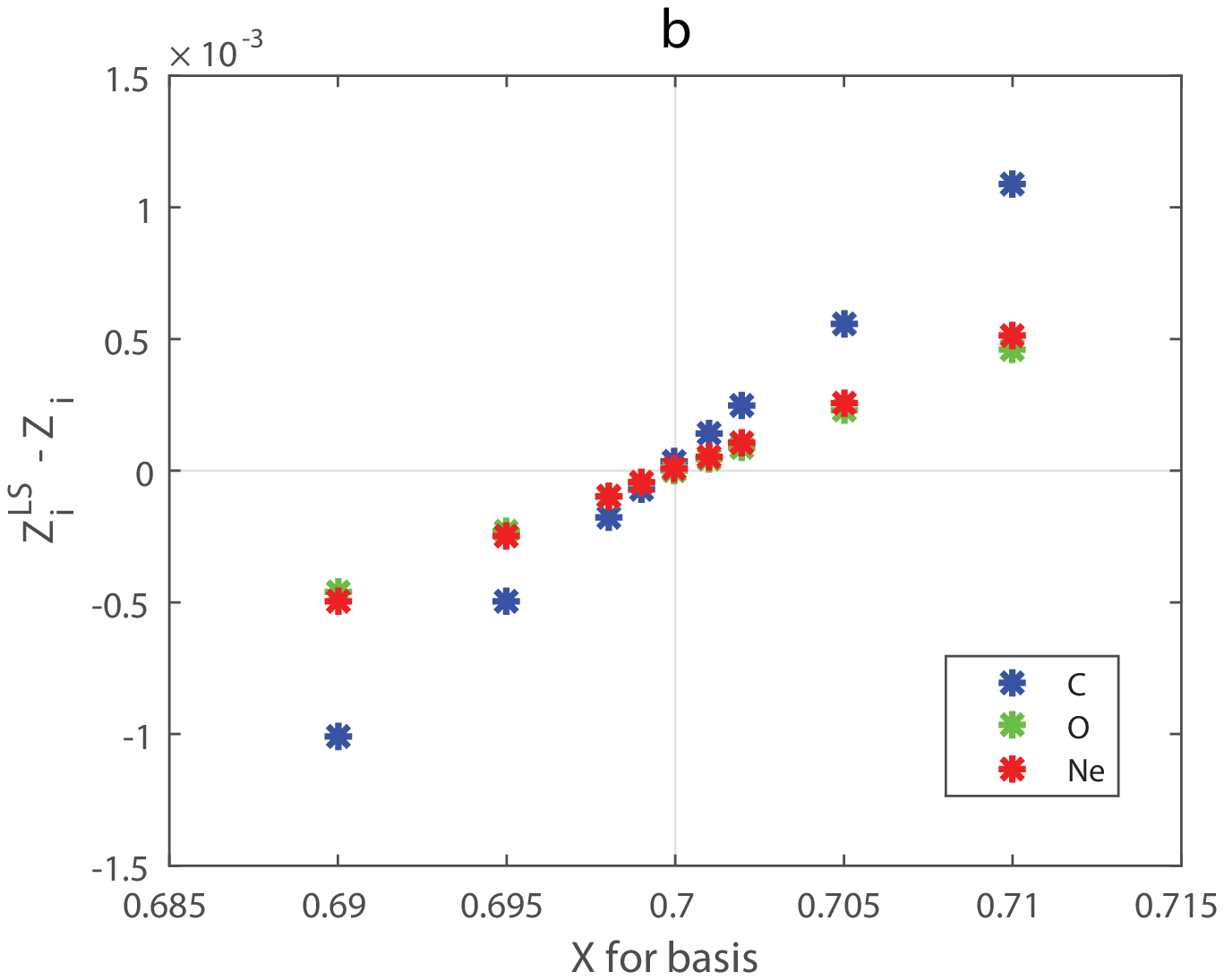}}
        \caption{(a) Standard deviation $\varepsilon $ obtained using LS method for various assumptions about hydrogen content $X$ . (b) Errors of mass fractions for carbon, oxygen, and neon at various $X$ in the background and basis.}
        \label{FigMeanDeviation}
\end{figure}

The problem of unknown $X$ can be resolved by varying hydrogen content in the background and basis and subsequently determining minimal $\varepsilon$. Figure~\ref{FigMeanDeviation}a presents the results of the approach. Standard deviation $\varepsilon$ is minimal at $X=0.70$ in the background and basis, which corresponds to the original value of hydrogen content in the analyzed profile $\Gamma_1$. Fig.~\ref{FigMeanDeviation}b shows the errors of LS estimations  at various $X$ in the background and basis, which are minimal at  $X=0.70.$

 This is a special experiment in which we assume that there are no errors, and only the parameter $X$ is unknown. The purpose of this epsilon analysis is not to obtain an independent $X$ determination, but the best $X$ value, for which the application of the LS decomposition is optimal. Using an independently provided $X$  that differs from our optimal value would lead to a deterioration of the LS estimation of the $Z$ abundances.

\section{Emulation of limited space resolution and pseudo-random and systematic errors}
\label{Sect_Errors}

We emulate random disturbances of $\Gamma_1$ at limited space resolution. The main reason for this is to obtain a relation between the amplitude of the disturbances and the level of error in the LS decomposition.

We consider an experiment with a $\Gamma_1$ profile that has some error function added to it. We choose an error function defined at 20 points with a spacing of $\Delta\lg T\approx 0.05$ in the $\lg T = 5.4 - 6.4$ range. At these points, random errors were generated according to a normal (Gaussian) distribution with standard deviation $\sigma_\Gamma=10^{-4}$.
Of course, this experiment is a particular one. However, we think that these added errors
are representative for possible errors in helioseismic inversions. The result is described in Appendix~\ref{Appendix_RandomErrors}, and it serves to estimate the error tolerance in the $\Gamma_1$ profile.

Obviously, adding errors to the $\Gamma_1$ profile deteriorates the results. However, if the standard deviation of errors $\sigma_{\Gamma}$ is under $10^{-4}$ , the abundance of oxygen could still be estimated with a relative accuracy of 10\%.

An additional experiment was performed with a set of $\Gamma_1$ points taken from a solar model based on the OPAL equation of state. Twenty points in radius were used. Details of this experiment are described in Appendix~\ref{Appendix_G1r}. This example of the decomposition somewhat differs from other experiments in the paper. The most represented elements (C, O, and Ne) are determined with a relative accuracy better than 10\%. Despite the satisfactory accuracy, we cannot draw any further conclusions, since this last experiment stands apart from the main idea of the paper and its interpretation would require considerable additional analysis.

\section{Conclusion}
\label{Sect_Conclusion}

The main result of the paper is that the $Z$ contribution in $\Gamma_1$ for a mixture of elements is a linear combination of the contributions of individual elements. Each contribution is proportional to the content of the element and does not depend on the contribution of other elements. The hypothesis of linearity is not always satisfied. In the following, we list the factors that are essential for linearity.

(i) If several elements simultaneously ionize, the interaction between them occurs through free electrons. The linearity of the contributions will be fulfilled in the ionization model of small admixtures at the background of almost completely ionized hydrogen and helium. Each heavy element is represented by a small mass fraction, and the ionization of hydrogen and helium practically does not change. Then, the total number of free electrons is approximately constant, which is a condition for the linearity of the contributions.

(ii) In addition to the linearity of the contributions of individual elements, the independence of the contributions from successive ionizations of one element plays an important role. If the ionization energies of the different ionization stages are close to each other, the contributions of the successive ionizations to $\Gamma_1$ are superimposed and summed up. This happens during the ionization of L electrons for example. As a result, the $Z$ contribution of the element becomes wide and poorly localized. A broad dip as a sum of narrow dips from sequential ionizations contains less information than several separated dips.  This circumstance complicates the use and interpretation of such contributions as basis functions. In this case, linearity is preserved, but sensitivity of $\Gamma_1$ to this element becomes weak due to poor localization of the ionization region by temperature.

(iii) On the other hand, the ionization of K electrons for C, N, O, and Ne plays an important role. The contribution to $\Gamma_1$ of K-electron ionization is relatively isolated from the contributions of other ionizations. Such contributions cover almost all the investigated adiabatic part of the convective zone, which allows us to use them as basis functions.

The inverse problem consists of estimating the content of elements based on a theoretical $\Gamma_1$ profile  and contributions from individual elements and $\Gamma_1$ for the H-He background.  We  use the LS method of decomposition of the $Z$ contribution $\Gamma_1$ by $Z$ contributions of eight elements. The method gives a satisfactory result in an  ideal numerical experiment with a known hydrogen content. Mass fractions of heavy elements are estimated with relative accuracy up to a few tenths of a percent.

Numerical experiments allow us to make several conclusions regarding LS estimations. There is a residual irreducible error in the LS method. It is caused by the nonlinearity of the $Z$ contribution for a mixture of elements. The error is marked $(\varepsilon) $ in
Fig.~\ref{Fig_LS_shema}. Our result consists of an estimation of the minimum value of this error in an ideal experiment.

 We now discuss the optimal number of elements in the basis. The contribution of additional elements in the basis turns out to be linearly dependent on the $Z$ contributions of the already included elements. This linear independence can be studied with an SVD decomposition of the array of the basis vectors.
Specifically, the condition number $\kappa = s_{max}/s_{min} $ for eight elements is $134$,
while for four major elements $\kappa = 10$. Here, $s_{max}$ and $s_{min}$ are the largest and smallest singular values. By adding additional elements to the basis, $s_{max}$ can only increase and $s_{min}$ can only decrease \citep{Golub_2013}. This proves that increasing the number of basis elements would not give a substantial gain but rather lead to an ill-posed LS problem.
Yet, a too-small number of elements would lead to an increase of $\varepsilon $, and consequently to large error in the abundance determination.
However, the precise optimal number of basic elements is unknown and will demand further study.

The standard deviation $\sigma $ (Eq.~(\ref{Eq_sigma})) characterizes the difference    between                                                                                                                             mass fractions $Z_{i}^{\mathrm{LS}}$ obtained using LS decomposition and original ${{Z}_{i}}$. The value of $\sigma $ is six-to-ten times greater than $\varepsilon $. It depends on the structure of the basis functions of the individual elements.

The analysis of $Z$ contributions depends on the hydrogen content $X$. The difference between the true $X$ and that used in the calculated basis contributions can lead to the appearance of an erroneous component in the profile ${{\delta }_{z}}{{\Gamma }_{1}}$.  However, we can perform computations for various $X$ and then use the value $X$ that gives minimal $\varepsilon $, and therefore we obtain an optimal result of the LS decomposition. This approach  is illustrated in Sect.~\ref{Subsect_UnknownX}.

The listed reasons do not cover all possible sources of errors.  For example, there is the possibility of an error in the assumed entropy. The results have an erroneous component if the basis functions are calculated for points of a different adiabat. It can be corrected using an approach such as that described for hydrogen, but a detailed experiment is beyond the scope of this paper.  Errors can also appear due to differences in the equations of state. Estimations of mass fractions can be unstable if the calculations are performed using different equations of state.

The basic conclusion is the existence of an ideal solution, with which it is possible to obtain a reliable estimation of the content of most abundant elements. The practical meaning of our results lies in finding the conditions under which the LS decomposition can give relevant results.

The LS estimation can be applied to a helioseismic inversion of $\Gamma_1$, provided that such a profile is obtained with a high accuracy. To successfully apply the method, the amplitude of error in the solar $\Gamma_1$ profile should be at least below $10^{-4}$. This is a necessary, but not sufficient condition.  Another important condition is that the theoretical equation of state should be close enough to the true plasma properties in the Sun. The measure of closeness can be expressed with the help of the epsilon deviation.

\begin{acknowledgements}

We owe sincere thanks to Prof. Douglas Gough whose inspiring idea motivated the present study. We are most grateful to Dr. Ga\"el Buldgen for helpful comments and discussions. The addition of the experiments in Section~3 (described in detail in Appendix B) was suggested by the anonymous referee. 

\end{acknowledgements}

\bibliographystyle{aa} 
\bibliography{Bibliography} 

\begin{thebibliography}{28}
\expandafter\ifx\csname natexlab\endcsname\relax\def\natexlab#1{#1}\fi

\bibitem[{{Antia} \& {Basu}(2006)}]{Antia_Basu_2006}
{Antia}, H.~M. \& {Basu}, S. 2006, \apj, 644, 1292

\bibitem[{{Asplund} {et~al.}(2021){Asplund}, {Amarsi}, \&
  {Grevesse}}]{Asplund_2021}
{Asplund}, M., {Amarsi}, A.~M., \& {Grevesse}, N. 2021, \aap, 653, A141

\bibitem[{{Asplund} {et~al.}(2005){Asplund}, {Grevesse}, \&
  {Sauval}}]{Asplund_2005}
{Asplund}, M., {Grevesse}, N., \& {Sauval}, A.~J. 2005, Astronomical Society of
  the Pacific Conference Series, Vol. 336, {The Solar Chemical Composition},
  ed. I.~{Barnes}, Thomas~G. \& F.~N. {Bash}, 25

\bibitem[{{Asplund} {et~al.}(2009){Asplund}, {Grevesse}, {Sauval}, \&
  {Scott}}]{Asplund_2009}
{Asplund}, M., {Grevesse}, N., {Sauval}, A.~J., \& {Scott}, P. 2009, \araa, 47,
  481

\bibitem[{{Ayukov} \& {Baturin}(2017)}]{Ayukov_Baturin_2017}
{Ayukov}, S.~V. \& {Baturin}, V.~A. 2017, Astronomy Reports, 61, 901

\bibitem[{{Baturin}(2010)}]{Baturin_2010}
{Baturin}, V.~A. 2010, \apss, 328, 147

\bibitem[{{Baturin} {et~al.}(2013){Baturin}, {Ayukov}, {Gryaznov},
  {Iosilevskiy}, {Fortov}, \& {Starostin}}]{Baturin_2013}
{Baturin}, V.~A., {Ayukov}, S.~V., {Gryaznov}, V.~K., {et~al.} 2013,
  Astronomical Society of the Pacific Conference Series, Vol. 479, {The Current
  Version of the SAHA-S Equation of State: Improvement and Perspective}, ed.
  H.~{Shibahashi} \& A.~E. {Lynas-Gray}, 11

\bibitem[{{Baturin} {et~al.}(2000){Baturin}, {D{\"a}ppen}, {Gough}, \&
  {Vorontsov}}]{Baturin_2000}
{Baturin}, V.~A., {D{\"a}ppen}, W., {Gough}, D.~O., \& {Vorontsov}, S.~V. 2000,
  \mnras, 316, 71

\bibitem[{{Baturin} {et~al.}(2017){Baturin}, {D{\"a}ppen}, {Morel}, {Oreshina},
  {Th{\'e}venin}, {Gryaznov}, {Iosilevskiy}, {Starostin}, \&
  {Fortov}}]{Baturin_2017}
{Baturin}, V.~A., {D{\"a}ppen}, W., {Morel}, P., {et~al.} 2017, \aap, 606, A129

\bibitem[{{Baturin} {et~al.}(2019){Baturin}, {D{\"a}ppen}, {Oreshina},
  {Ayukov}, \& {Gorshkov}}]{Baturin_2019}
{Baturin}, V.~A., {D{\"a}ppen}, W., {Oreshina}, A.~V., {Ayukov}, S.~V., \&
  {Gorshkov}, A.~B. 2019, \aap, 626, A108

\bibitem[{{Buldgen} {et~al.}(2017){Buldgen}, {Salmon}, {Noels}, {Scuflaire},
  {Dupret}, \& {Reese}}]{Buldgen_2017}
{Buldgen}, G., {Salmon}, S.~J.~A.~J., {Noels}, A., {et~al.} 2017, \mnras, 472,
  751

\bibitem[{{Cox} \& {Giuli}(1968)}]{Cox_Giuli_1968}
{Cox}, J.~P. \& {Giuli}, R.~T. 1968, {Principles of stellar structure} (Gordon
  and Breach, New York)

\bibitem[{{D\"appen} \& {Gough}(1984)}]{Dappen_Gough_1984}
{D\"appen}, W. \& {Gough}, D.~O. 1984, in Liege International Astrophysical
  Colloquia, Vol.~25, Liege International Astrophysical Colloquia, 264--268

\bibitem[{{D{\"a}ppen} {et~al.}(1993){D{\"a}ppen}, {Gough}, {Kosovichev}, \&
  {Rhodes}}]{Dappen_1993}
{D{\"a}ppen}, W., {Gough}, D.~O., {Kosovichev}, A.~G., \& {Rhodes}, E.~J., J.
  1993, Astronomical Society of the Pacific Conference Series, Vol.~40, {On the
  Influence of Treatment of Heavy Elements in the Equation of State on the
  Resulting Values of the Adiabatic Exponent}, ed. W.~W. {Weiss} \&
  A.~{Baglin}, 304

\bibitem[{{Elliott}(1996)}]{Elliott_1996}
{Elliott}, J.~R. 1996, \mnras, 280, 1244

\bibitem[{{Golub} \& {Van Loan}(2013)}]{Golub_2013}
{Golub}, G.~H. \& {Van Loan}, C.~F. 2013, {Matrix Computations}, 4th edn.
  (Johns Hopkins University Press, Baltimore)

\bibitem[{{Gong} {et~al.}(2001){Gong}, {D{\"a}ppen}, \& {Nayfonov}}]{Gong_2001}
{Gong}, Z., {D{\"a}ppen}, W., \& {Nayfonov}, A. 2001, \apj, 563, 419

\bibitem[{{Grevesse} \& {Noels}(1993)}]{Grevesse_Noels_1993}
{Grevesse}, N. \& {Noels}, A. 1993, in Origin and Evolution of the Elements,
  ed. N.~{Prantzos}, E.~{Vangioni-Flam}, \& M.~{Casse}, 15--25

\bibitem[{{Gryaznov} {et~al.}(2006){Gryaznov}, {Ayukov}, {Baturin},
  {Iosilevskiy}, {Starostin}, \& {Fortov}}]{Gryaznov_2006}
{Gryaznov}, V.~K., {Ayukov}, S.~V., {Baturin}, V.~A., {et~al.} 2006, Journal of
  Physics A Mathematical General, 39, 4459

\bibitem[{{Gryaznov} {et~al.}(2013){Gryaznov}, {Iosilevskiy}, {Fortov},
  {Starostin}, {Roerich}, {Baturin}, \& {Ayukov}}]{Gryaznov_2013}
{Gryaznov}, V.~K., {Iosilevskiy}, I.~L., {Fortov}, V.~E., {et~al.} 2013,
  Contributions to Plasma Physics, 53, 392

\bibitem[{{Hansen} {et~al.}(2004){Hansen}, {Kawaler}, \&
  {Trimble}}]{Hansen_2004}
{Hansen}, C.~J., {Kawaler}, S.~D., \& {Trimble}, V. 2004, {Stellar interiors :
  physical principles, structure, and evolution} (Springer-Verlag, New York)

\bibitem[{{Kosovichev} {et~al.}(1992){Kosovichev}, {Christensen-Dalsgaard},
  {D\"appen}, {Dziembowski}, {Gough}, \& {Thompson}}]{Kosovichev_1992}
{Kosovichev}, A.~G., {Christensen-Dalsgaard}, J., {D\"appen}, W., {et~al.}
  1992, \mnras, 259, 536

\bibitem[{{Lin} {et~al.}(2007){Lin}, {Antia}, \& {Basu}}]{Lin_2007}
{Lin}, C.-H., {Antia}, H.~M., \& {Basu}, S. 2007, \apj, 668, 603

\bibitem[{{Lin} \& {D{\"a}ppen}(2005)}]{Lin_Dappen_2005}
{Lin}, C.-H. \& {D{\"a}ppen}, W. 2005, \apj, 623, 556

\bibitem[{{Rogers} \& {Nayfonov}(2002)}]{Rogers_Nayfonov_2002}
{Rogers}, F.~J. \& {Nayfonov}, A. 2002, \apj, 576, 1064

\bibitem[{{Rogers} {et~al.}(1996){Rogers}, {Swenson}, \&
  {Iglesias}}]{Rogers_1996}
{Rogers}, F.~J., {Swenson}, F.~J., \& {Iglesias}, C.~A. 1996, \apj, 456, 902

\bibitem[{{Vorontsov} {et~al.}(2013){Vorontsov}, {Baturin}, {Ayukov}, \&
  {Gryaznov}}]{Vorontsov_2013}
{Vorontsov}, S.~V., {Baturin}, V.~A., {Ayukov}, S.~V., \& {Gryaznov}, V.~K.
  2013, \mnras, 430, 1636

\bibitem[{{Vorontsov} {et~al.}(2014){Vorontsov}, {Baturin}, {Ayukov}, \&
  {Gryaznov}}]{Vorontsov_2014}
{Vorontsov}, S.~V., {Baturin}, V.~A., {Ayukov}, S.~V., \& {Gryaznov}, V.~K.
  2014, \mnras, 441, 3296

\end{thebibliography}


\begin{appendix}
        
\section{Relation between ionization of elements and their contributions to $\delta_Z\Gamma_1$ }
\label{Appendix_Ionization}

This appendix presents a simple picture of the plasma elasticity decrease when a heavy element is ionized. The definition of the adiabatic exponent ${\Gamma _1} = {\left. {\left( {\partial \ln P/\partial \ln \rho } \right)} \right|_S}$, with the help of the basic thermodynamic law for adiabatic processes $dU =  - PdV,$  can be rewritten as
\begin{equation}
{\Gamma _1} = {\left. {\frac{{\partial ({P \mathord{\left/
                                                {\vphantom {P \rho }} \right.
                                                \kern-\nulldelimiterspace} \rho })}}{{\partial U}}} \right|_S} + 1.
                                        \label{Eq_G1}
\end{equation}

\noindent
In an ideal plasma, without the ionization, the internal energy and pressure are given by
\begin{equation}
U = \frac{3}{2}N{k_B}T\quad \mathrm{ and }\quad P/\rho  = N{k_B}T,
\end{equation}

\noindent
where $N$ is total number of all particles (atoms, ions, and electrons) per gram, and ${k_B}$ is theBoltzmann constant. So in this case $U=\left(3/2\right)P/\rho$ and from Eq.~(\ref{Eq_G1}) one has ${\Gamma _1} = 5/3.$

Ionization increases the internal energy through ionization energy ${U_{\mathrm{ioniz}}}$ \citep{Hansen_2004}:
\begin{equation}
U = \frac{3}{2}N{k_B}T + {U_{\mathrm{ioniz}}}.
\end{equation}  

\noindent
Adiabatic exponent ${\Gamma _1}$ changes in the presence of ionization. Equation~(\ref{Eq_G1}) can be rewritten in the following form:
\begin{equation}
{\Gamma _1} = \frac{1}{{{{\left. {\frac{{\partial U}}{{\partial \left( {P/\rho } \right)}}} \right|}_S}}} + 1 = \frac{1}{{\frac{3}{2} + {{\left. {\frac{{\partial {U_{\mathrm{ioniz}}}}}{{\partial \left( {P/\rho } \right)}}} \right|}_S}}} + 1 = \frac{2}{3}\frac{1}{{1 + \frac{2}{3}{{\left. {\frac{{\partial {U_{\mathrm{ioniz}}}}}{{\partial \left( {P/\rho } \right)}}} \right|}_S}}} + 1.
\label{Eq_G1_ioniz}
\end{equation}  

\noindent
The derivative in the denominator is smaller than unity in the case of ionization of a small abundant
element, and expression \ref{Eq_G1_ioniz} is expanded into a Taylor series:
\begin{equation}
{\Gamma _1} \approx \frac{2}{3}\left( {1 - \frac{2}{3}{{\left. {\frac{{\partial {U_{\mathrm{ioniz}}}}}{{\partial \left( {P/\rho } \right)}}} \right|}_S}} \right) + 1 = \frac{5}{3} - \frac{4}{9}{\left. {\frac{{\partial {U_{\mathrm{ioniz}}}}}{{\partial \left( {P/\rho } \right)}}} \right|_S}.
\end{equation}

\noindent
The derivative in the second term is positive in ionization regions, and therefore ${\Gamma _1}$ becomes smaller.

As an example, we apply these expressions to small impurity of carbon at the background of fully ionized
hydrogen and helium. Mass percentage of carbon is 0.01, hydrogen 0.70, helium 0.29. The sequence  $(T,\rho)$  is taken from the model adiabatic curve corresponding to deep part of the convection zone. Therefore, the analytical description can be applied.

A carbon atom has six electrons: four on the outer L-shell and two on the inner K-shell. The ionization internal energy is written in two forms by regrouping the sum:
\begin{equation}
{U_{\mathrm{ioniz}}} = \sum\limits_{j = 1}^6 {\left( {{N_{{C^{j}}}}\sum\limits_{i = 1}^j {{\chi _{{C^{i}}}}} } \right) = } \sum\limits_{j = 1}^6 {\left( {{\chi _{{C^{j }}}}\sum\limits_{i = j}^6 {{N_{{C^{i}}}}} } \right).}
\label{Eq_Uioniz}
\end{equation}

\begin{figure}
        \centering
        \resizebox{\hsize}{!}{\includegraphics{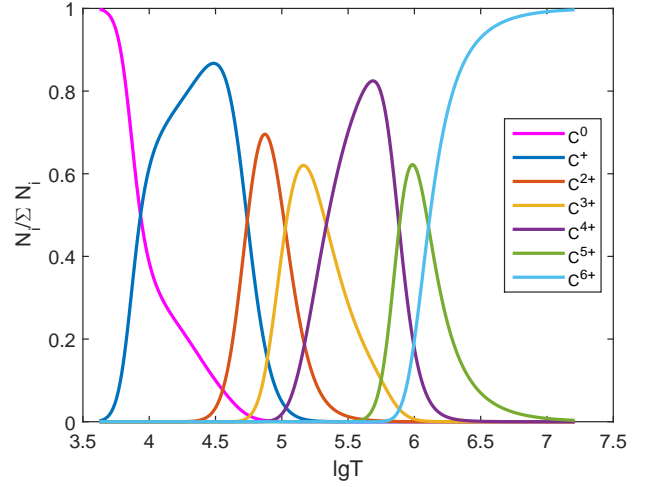}}
        \caption{Distribution of carbon atoms and ions along the adiabatic curve.}
        \label{Fig_NC_lgT}
\end{figure}

\noindent
Here, $\chi_{{C^{i}}}$ is the ionization potential of  the j-th ion of carbon, and ${N_{{C^{j}}}}$ represents the concentration of the j-th ion. The concentration of ions along the selected adiabatic curve is shown in Fig.~\ref{Fig_NC_lgT}. Carbon is mainly in the neutral state at low temperatures $\lg T < 3.7$, and  it is almost fully ionized at $\lg T > 6.7$.

\begin{figure}
        \centering
        \resizebox{\hsize}{!}{\includegraphics{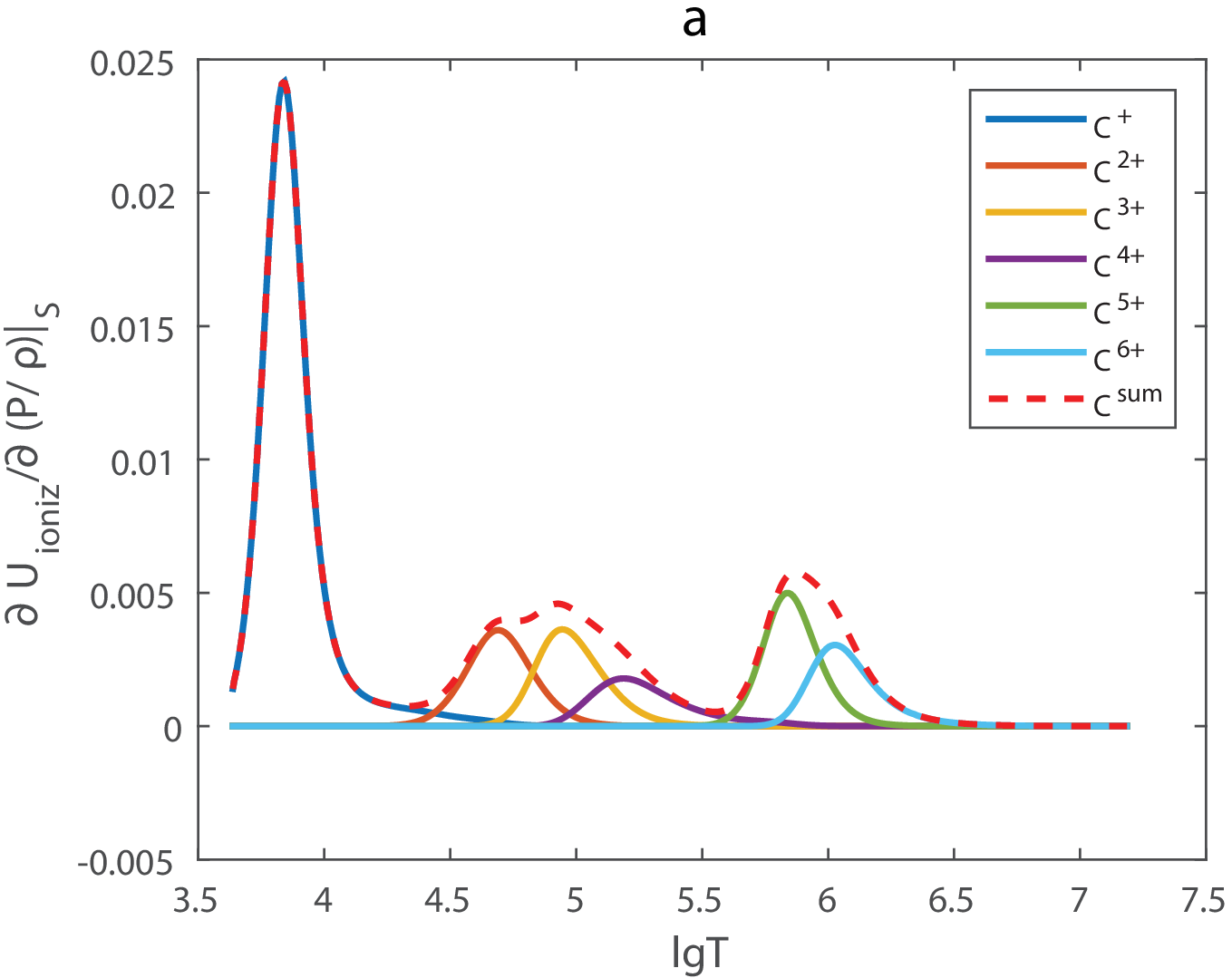}}
        \resizebox{\hsize}{!}{\includegraphics{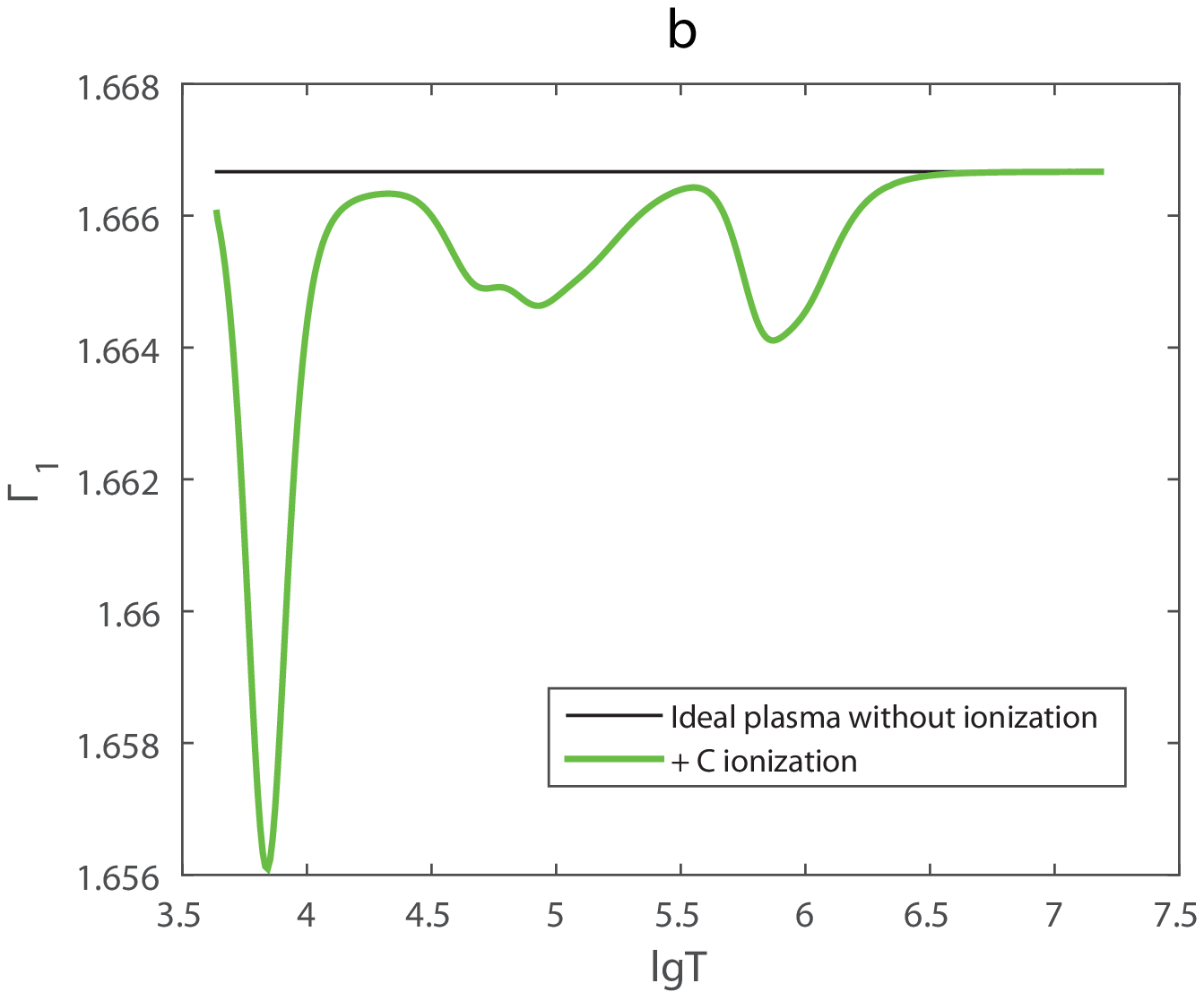}}
        \caption{(a)  Derivative ${{\left. \left[ \partial {{U}_{\mathrm{ioniz}}}/\partial \left( P/\rho  \right) \right] \right|}_{S}}$ for the case of carbon ionization (red dashed curve), and derivatives of terms ${{\chi }_{{{C}^{j}}}}\sum\limits_{i=j}^{6}{{{N}_{{{C}^{i}}}}}$(colored solid curves). (b) Adiabatic exponent $\Gamma_1$ for ideal plasma without ionization (black horizontal line) and with ionization (green curve).}
        \label{Fig_dU}
\end{figure}

Derivative ${\left. {\left[ {\partial {U_{\mathrm{ioniz}}}/\partial \left( {P/\rho } \right)} \right]} \right|_S}$ is shown in Fig.~\ref{Fig_dU}a by the red dashed curve together with the individual
terms of expression (\ref{Eq_Uioniz}). The dotted line has three local maxima. The first peak, at $\lg T = 3.8$, is due to the ionization of neutral atoms. The second one, at $\lg T = 4.6 - 5$, is associated with the ionization of the L shell. The third maximum, at $\lg T = 5.9$, corresponds to the ionization of the K shell.  We note that individual ionizations are linearly added in total derivatives in our approximation. Outside of the carbon ionization range, the derivative goes to zero. The ${\Gamma _1}$-lowering is shown in Fig.~\ref{Fig_dU}b. At the temperatures $\lg T = 5.4 - 6.4$, which are analyzed in the article, the
${\Gamma _1}$-lowering is mainly due to the ionization of the carbon K-shell electron.       

Simple considerations in this appendix explain the formation of the ${\Gamma _1}$ lowering of each element as a sum of
individual contributions of every ionization stage. Position of minima in the ${\Gamma _1}$-lowering profile is directly associated with the ionization energy of corresponding ions.
For computations that are more accurate, the Coulomb interaction, electron degeneracy, and
simultaneous background ionization should be taken into account. All of them are included in the SAHA-S
calculation used in our work.

\section{Some examples of error effect}
\label{Appendix_Errors}

\subsection{Background adiabatic exponent for hydrogen-helium plasma }
\label{Appendix_G1_HHe}

\begin{figure}
        \centering
        \resizebox{\hsize}{!}{\includegraphics{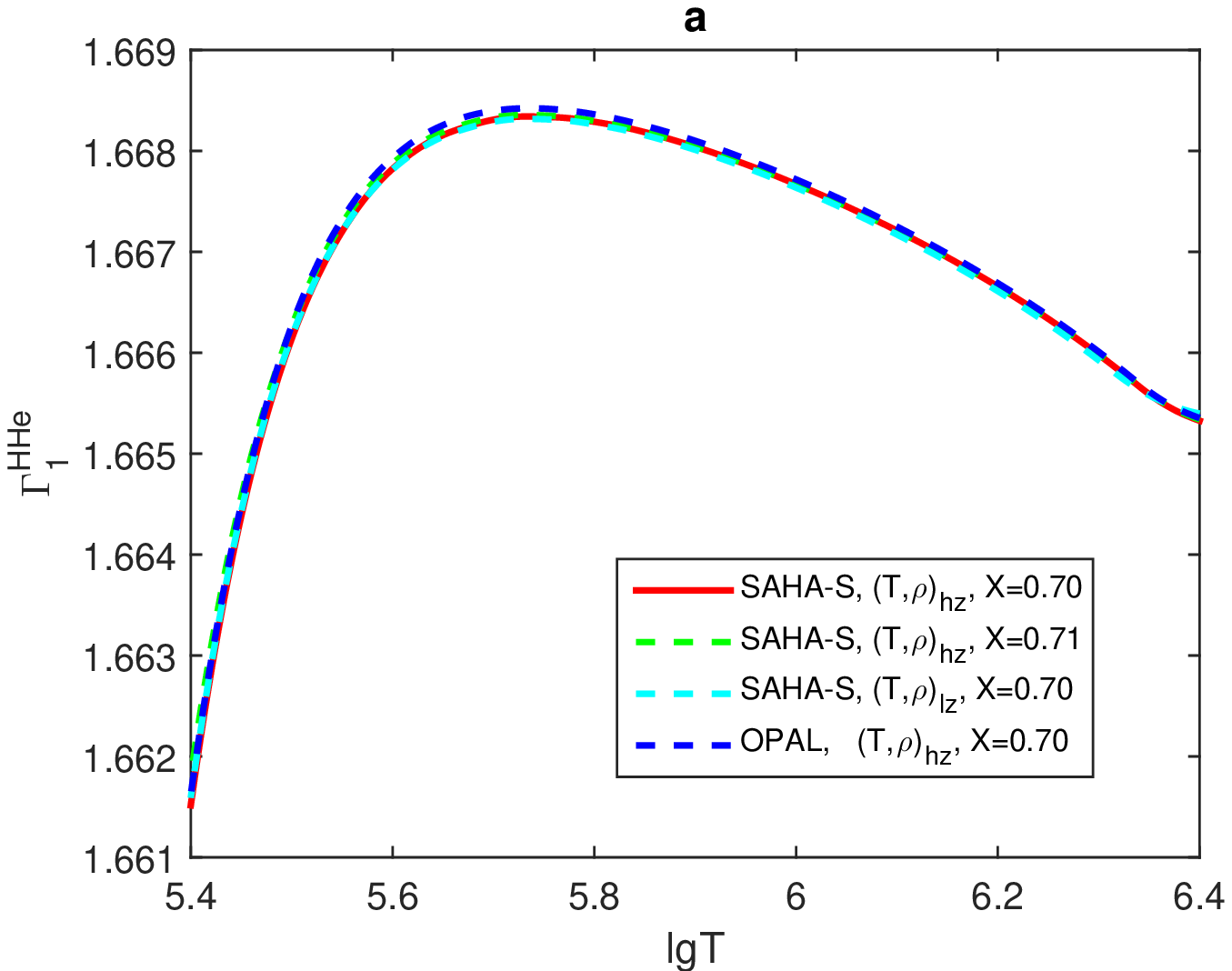}}
        \resizebox{\hsize}{!}{\includegraphics{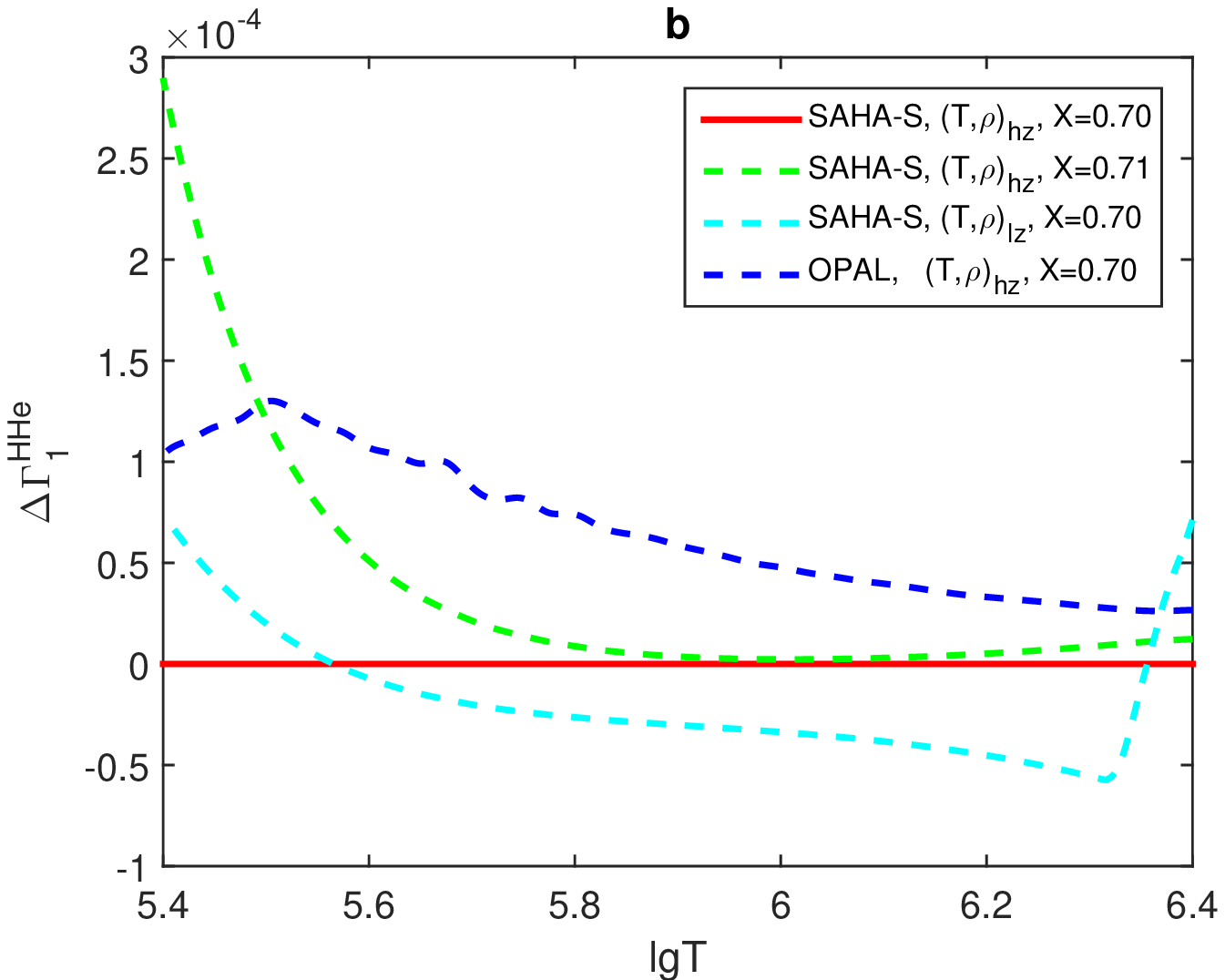}}
        \caption{(a) Adiabatic exponents for hydrogen-helium plasma for different input parameters. (b)  Difference between them.}
        \label{Fig_G1_HHe}
\end{figure}

We used the adiabatic exponent $\Gamma _{1}^{\mathrm{HHe}}$ to compute the $Z$ contributions. It depends on the hydrogen mass fraction $X$, the $\left( T,\rho  \right)$ profile that is on entropy, and the equation of state.  Figure~\ref{Fig_G1_HHe}a shows several examples of the $\Gamma _{1}^{\mathrm{HHe}}$ profiles for different parameters. Mass fractions of hydrogen are $X=0.70$ and $X=0.71$. $\left( T,\rho  \right)$ profiles are taken from the standard solar models with high and low mass fractions of heavy elements Z (corresponding to models 771-0001 and 771-0002 from \citep{Ayukov_Baturin_2017}). There are two examples of ${{\Gamma }_{1}}$ profiles computed with the SAHA-S and OPAL equations of state. The adiabatic exponents are very similar on the figure scale. The detailed difference between them is presented in Fig.~\ref{Fig_G1_HHe}b.

The red curve represents $\Gamma _{1}^{\mathrm{HHe}}$ computed for $X=0.70$, at points $\left( T,\rho  \right)$ from the high-$Z$ model and with the SAHA-S equation of state. The green curve differs from it only by hydrogen, which is $X=0.71$.
The cyan curve corresponds to $\Gamma _{1}^{\mathrm{HHe}}$ computed with the same $X$, but on points $\left( T,\rho  \right)$ from the low-$Z$ model. The deviation at high temperatures is due to the shallow convection zone in the standard low-$Z$ model.
The blue curve shows effect of the equation of state. There, the OPAL is used instead of SAHA-S. The differences are rather moderate and do not exceed $1.3\times {{10}^{-4}}$. The intrinsic distinction between equations of state demonstrates different behavior in case $X$ or adiabat variations.
All these curves demonstrate the behavior of the $\Gamma _{1}^{\mathrm{HHe}}$ profile itself, and these variations should not be translated to errors in ${{\Gamma }_{1}}$ contribution.
On the contrary, this effect can be used as diagnostic tool.

\subsection{LS decomposition for different $(T,\rho)$ profiles   }
\label{Appendix_TRho}

\begin{figure}
        \centering
        \resizebox{\hsize}{!}{\includegraphics{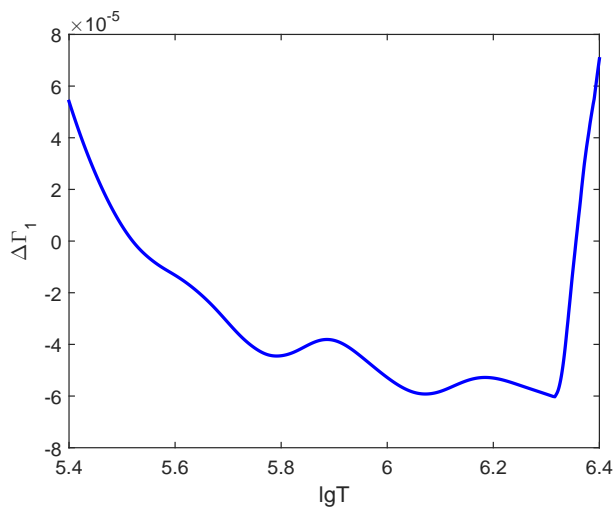}}
        \caption{Difference between $\Gamma_1$ profiles computed at points $\left( T,\rho  \right)$ from low-$Z$ and high-$Z$ models. Computations are performed with SAHA-S for the AGSS09 mixture, $X=0.70$, $Z=0.02$.}
        \label{Fig_dG1_TRho}
\end{figure}

In the main part of the paper, we considered cases when points $\left( T,\rho  \right)$ were taken from the high-$Z$ model 771-0001 to compute the analyzed $\Gamma_{1}$, background $\Gamma _{1}^{\mathrm{HHe}}$, and basis $\delta _{Z}^{i}{{\Gamma }_{1}}$.
Let us suppose that we do not know the points $\left( T,\rho  \right)$ of the analyzed profile  ${{\Gamma }_{1}}$. We take the profile ${\left( T,\rho  \right)}_{\mathrm{hz}}$ from the high-$Z$ model 771-0001 to compute $\Gamma _{1}^{\mathrm{HHe}}$ and $\delta _{Z}^{i}{{\Gamma }_{1,}}$ whereas  the analyzed ${{\Gamma }_{1}}$ is computed for the ${\left( T,\rho  \right)}_{\mathrm{lz}}$ from the low-$Z$ model 771-0002. As in the main part, $X=0.70$ and $Z=0.02$ for the mixture AGSS09.  Figure~\ref{Fig_dG1_TRho} presents the difference between ${{\Gamma }_{1}}$ profiles computed on points $\left( T,\rho  \right)$ from low-$Z$ and high-$Z$ models. The difference does not exceed $8\times {{10}^{-5}}$. Thus, we study how this difference influences the results of LS-decomposition.

The results are presented in Table~\ref{Table_dG1_TRho}. The mean deviation is $\varepsilon =2.849\times {{10}^{-6}}$, which is an order of magnitude larger than in the ideal case for which the profiles $\left( T,\rho  \right)$ coincide (see Table~\ref{Table_LS_Decomposition} of the main part). The mass fractions of the most abundant elements C, N, O, and Ne are determined with an accuracy of a few percent.   Thus, an error in the $\left( T,\rho  \right)$ profile leads to increasing $\varepsilon $. This property can be used for diagnostic purposes.

\begin{table}
        \caption{LS decomposition for different $(T,\rho)$  profiles in the analyzed function and background.}
        \centering
        \begin{tabular}{|l|c|c|}
                \hline
                Elem. & $Z_i^{\mathrm{LS}}, \times 10^{-2} $    & $\Delta Z_i/Z_i$, \% \\
                \hline  
                C &     0.3724  & 2.5 \\
                N &     0.1026  &-3.6 \\
                O &     0.8876  & 0.8 \\
                Ne&     0.1959  & 1.5 \\
                Mg&     0.0969  &-10.9 \\
                Si&     0.1244  & 21.8 \\
                S &     0.0361  &-24.0 \\
                Fe&     0.1947  &-1.9 \\
                \hline
                $Z^{\mathrm{LS}}$ & \multicolumn{2}{|c|}{0.020105} \\
                \hline  
                $\eta$   & \multicolumn{2}{|c|}{$1.445\times 10^{-5}$}   \\
                \hline  
                $\varepsilon$ & \multicolumn{2}{|c|}{$2.849\times 10^{-6}$}   \\
                \hline  
                $\sigma$ & \multicolumn{2}{|c|}{$1.417\times 10^{-5}$}   \\
                \hline
        \end{tabular}
        \label{Table_dG1_TRho}
\end{table}

\subsection{LS decomposition for the OPAL equation of state}
\label{Appendix_OPALeos}

\begin{figure}
        \centering
        \resizebox{\hsize}{!}{\includegraphics{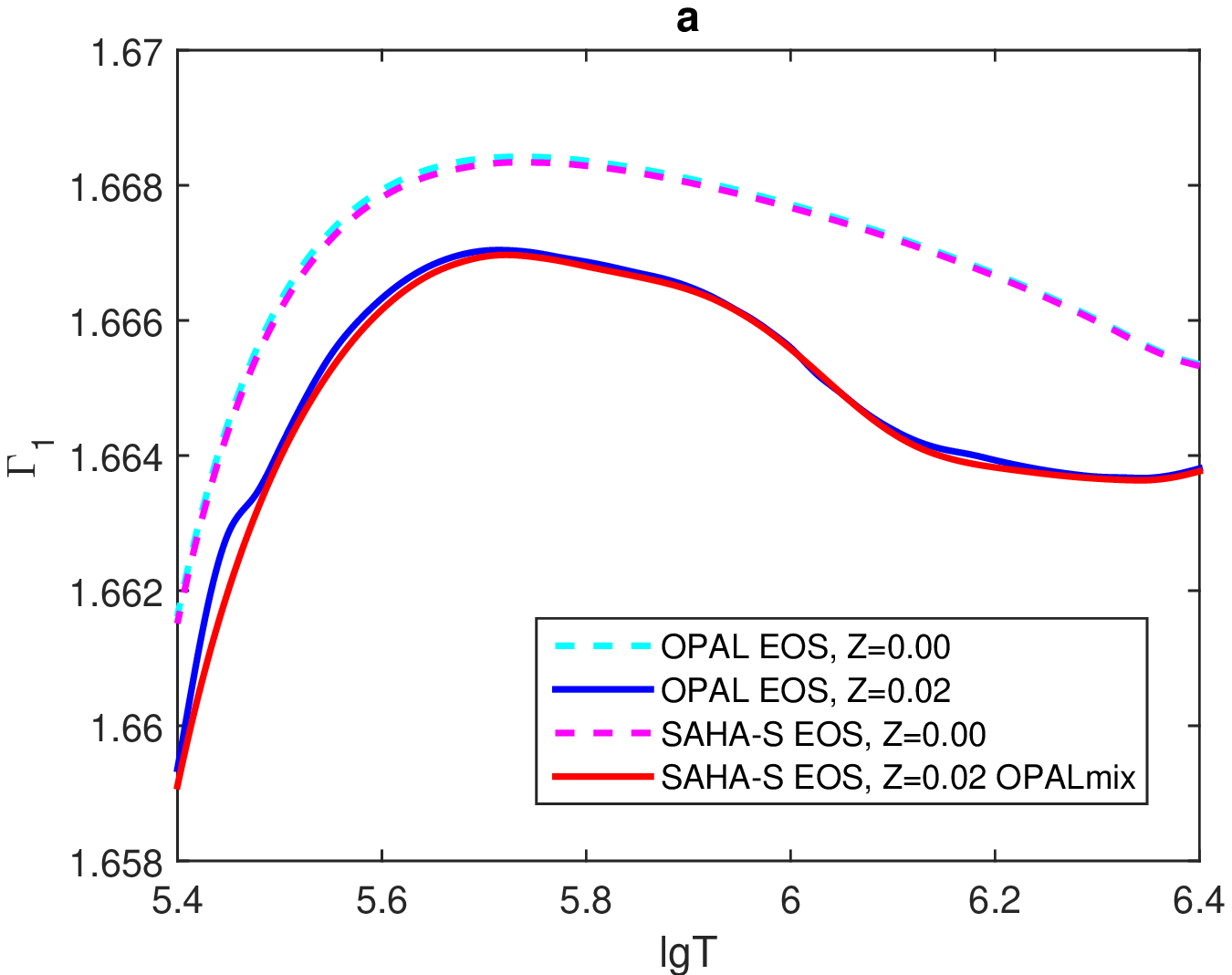}}
        \resizebox{\hsize}{!}{\includegraphics{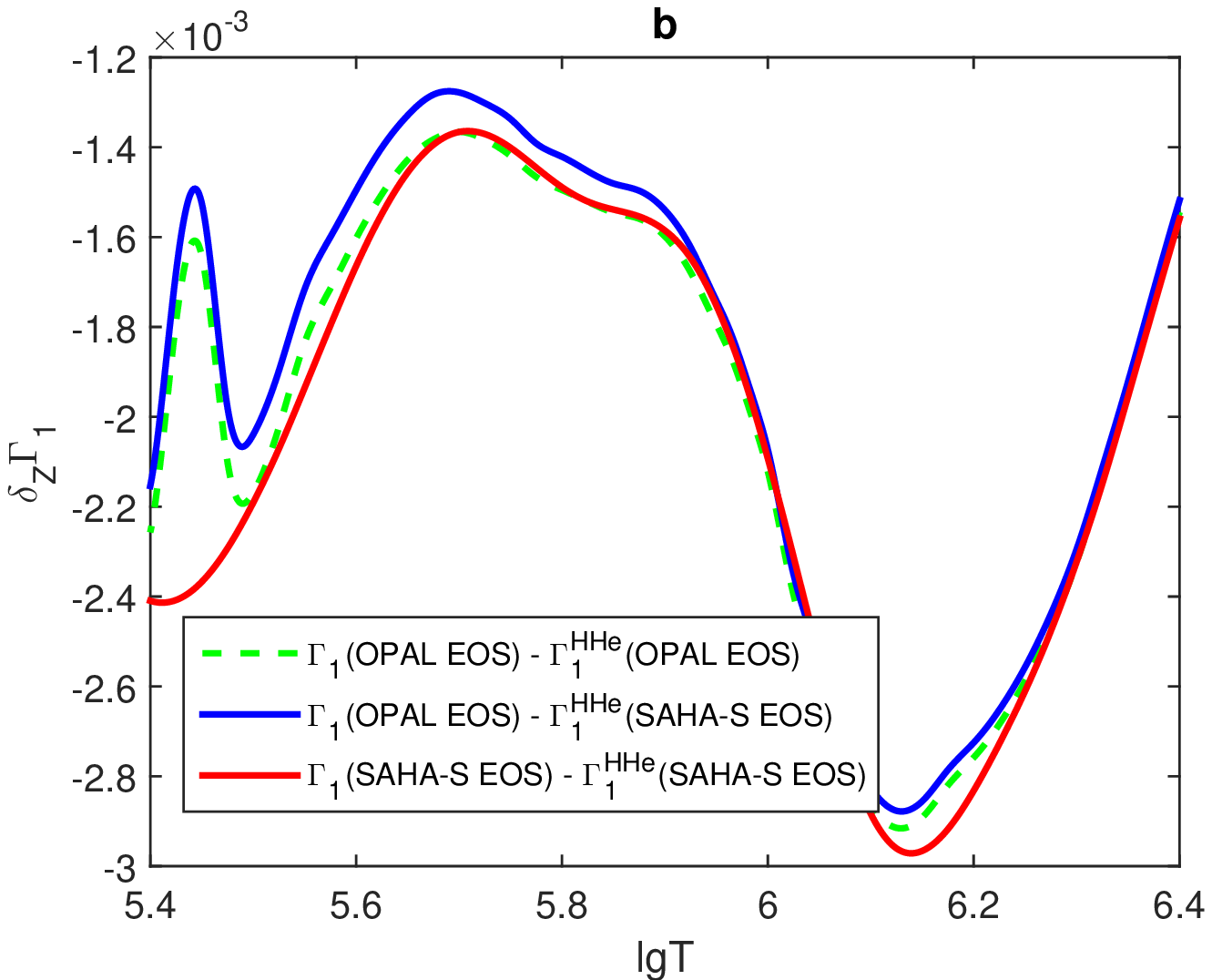}}
        \caption{(a) Adiabatic exponents and (b) $Z$ contributions  computed in the OPAL and SAHA-S (OPAL mixture) equations of state.
         $X=0.70$, $Z=0.02$.}
        \label{Fig_OPALeos_OPALmix}
\end{figure}

Figure~\ref{Fig_OPALeos_OPALmix}a shows adiabatic exponents computed in the OPAL and SAHA-S (OPAL mixture)
equations of state  at points $(T_j,\rho_j)$ from the solar model 771-0001. The curves for a hydrogen-helium plasma (dashed lines) agree well,
whereas curves for $Z=0.02$ (solid lines) differ by about $10^{-3}$ at $\lg T=5.4-5.5$
because of spurious fluctuations contained in OPAL.

Figure~\ref{Fig_OPALeos_OPALmix}b shows $Z$ contributions computed in different ways. The peak at low temperature region $\lg T=5.4-5.5$  is caused by the artefact in $\Gamma_{1}$ at $Z=0.02$ in OPAL.

The $Z$ contribution for OPAL was calculated in two ways. In the first case, the $\delta_Z\Gamma_1^{\mathrm{HHe}}$ for the hydrogen-helium plasma was calculated with the SAHA-S (blue curve), in the second case it was calculated with OPAL (green dashed curve). The difference between the two values of $\delta _Z\Gamma_1$  is about $10^{-4}$.
The origin of this difference is unknown and will require additional research. These examples are given only to demonstrate the required level of precision of a modern equation of state and the absolute accuracy with which the solar equation of state has to be known.

Table~\ref{Table_OPALeos} shows the results of the LS method for the $Z$ contribution from OPAL. Decomposition over the entire interval $\lg T=5.4-6.4$ gives an accuracy of  14\% for the oxygen mass fraction and 81\% for carbon. A bad carbon estimation is caused by a peak in $\delta_Z\Gamma_1$ at low temperatures. If the investigated temperature range is reduced to $\lg T=5.5-6.4$ and thereby the peak is removed, then the estimates are improved: oxygen 8\%, carbon 14\%. Switching the profile $\Gamma_1^{\mathrm{HHe}}$ from SAHA-S to OPAL is not essential.

\begin{table*}
        \caption{LS decomposition for OPAL equation of state calculated at different intervals $\lg T$ and with different background $\Gamma _{1}^{\mathrm{HHe}}$ .}
        \centering
        \begin{tabular}{|l| *{3}{c|c|} }
                \hline
                & \multicolumn{2}{c|}{$\Gamma_1^{\mathrm{HHe}}$ from SAHA-S, }
                & \multicolumn{2}{c|}{$\Gamma_1^{\mathrm{HHe}}$ from SAHA-S, }
                & \multicolumn{2}{c|}{$\Gamma_1^{\mathrm{HHe}}$ from OPAL, }  \\
                & \multicolumn{2}{c|}{$\lg T=5.4-6.4$}
                & \multicolumn{2}{c|}{$\lg T=5.5-6.4$}
                & \multicolumn{2}{c|}{$\lg T=5.5-6.4$}  \\              
                \cline{2-7}
                Elem.& $Z_i^{\mathrm{LS}},$ $\times 10^{-2}$  &  ${\Delta Z_i}/{Z_i},$ \%
                &   $Z_i^{\mathrm{LS}},$ $\times 10^{-2}$   &  ${\Delta Z_i}/{Z_i},$ \%
                &  $Z_i^{\mathrm{LS}},$ $\times 10^{-2}$   & ${\Delta Z_i}/{Z_i},$  \%  \\
                \hline
                C  & 0.0738 & -81 &  0.4262 & 12 &  0.4301 &  13 \\
                N  & 0.0763 & -32 &  0.1017 & -9 &  0.1025 &  -8 \\
                O  & 0.9288 & -14 &  0.9993 & -8 &  1.0187 &  -6 \\
                Ne & 0.3522 & -16 &  0.5298 & 26 &  0.5380 &  28 \\
                Mg &-0.8653 &  -  & -0.2545 & -  & -0.2295 &  -  \\
                Si & 0.7998 &  -  &  0.1474 & -  &  0.1814 &  -  \\
                S  &-0.0370 &  -  & -0.0112 & -  & -0.0387 &  -  \\
                Fe & 0.3221 &  -  & -0.0308 & -  & -0.0370 &  -  \\
                \hline
                $Z^{\mathrm{LS}}$& \multicolumn{2}{c|}{0.016089}  & \multicolumn{2}{c|}{0.019079} & \multicolumn{2}{c|}{0.019557}  \\
                \hline
                $\eta$& \multicolumn{2}{c|}{$1.788\times 10^{-4}$}  & \multicolumn{2}{c|}{$8.683\times 10^{-5}$} & \multicolumn{2}{c|}{$4.326\times 10^{-5}$}  \\
                \hline
                $\varepsilon$& \multicolumn{2}{c|}{$7.107\times 10^{-5}$} & \multicolumn{2}{c|}{$1.950\times 10^{-5}$} & \multicolumn{2}{c|}{$1.824\times 10^{-5}$}\\
                \hline
                $\sigma$ & \multicolumn{2}{c|}{$1.641\times 10^{-4}$} & \multicolumn{2}{c|}{$8.461\times 10^{-5}$} & \multicolumn{2}{c|}{$3.922\times 10^{-5}$} \\
                \hline
        \end{tabular}
        \label{Table_OPALeos}
\end{table*}

The analysis of the OPAL equation of state thus makes it possible to determine oxygen, as the most abundant heavy element, with an accuracy of about 10\%. Selecting a background $\Gamma_1^{\mathrm{HHe}}$ from SAHA-S or OPAL is not essential. An accurate equation of state plays an important role in the reliability of our method.

\subsection{An example of finite space resolution and random errors}
\label{Appendix_RandomErrors}

Until now, all computations were performed for 1309 points $\left( T_j,\rho_j \right)$ in the interval $\lg T=5.4-6.4$, i.e. simulating a continuous function $\Gamma_1$. Generally, such a function can show some small details that cannot be expected in helioseismic profiles $\Gamma_1$. The expected resolution for the inverted profile $\Gamma_1$ is not much better than $\Delta \lg T\approx 0.05$.

To simulate a profile with a finite resolution, we switched to a discrete version of LS decomposition with a limited number ND of points. Varying the ND from 1309 to about 40-20, we did not find significant changes in the LS decomposition results. In Table~\ref{Table_20points}, we present the results for the discrete case with ND = 20. It can be compared with Table~\ref{Table_LS_Decomposition} (AGSS09). Thus, the discrete LS decomposition is computationally equivalent to its quasi-continuous counterpart in the case of an ideal experiment.

\begin{table}
        \caption{LS decomposition for $\Gamma_1$ profile with limited space resolution (20 points instead of 1309).}
        \centering
        \begin{tabular}{|l|c|c|}
                \hline
                Elem. & $Z_i^{\mathrm{LS}}, \times 10^{-2} $    & $\Delta Z_i/Z_i$, \% \\
                \hline  
                C &     0.3662  & 0.8 \\
                N &     0.1064  &-0.02 \\
                O &     0.8803  &-0.02 \\
                Ne&     0.1936  & 0.3 \\
                Mg&     0.1137  & 4.6 \\
                Si&     0.0988  &-3.2 \\
                S &     0.0468  &-1.5 \\
                Fe&     0.1965  &-0.9 \\
                \hline
                $Z^{\mathrm{LS}}$ & \multicolumn{2}{|c|}{0.020024} \\
                \hline  
                $\eta$   & \multicolumn{2}{|c|}{$2.232\times 10^{-6}$}   \\
                \hline  
                $\varepsilon$ & \multicolumn{2}{|c|}{$2.336\times 10^{-7}$}   \\
                \hline  
                $\sigma$ & \multicolumn{2}{|c|}{$2.219\times 10^{-6}$}   \\
                \hline
        \end{tabular}
        \label{Table_20points}
\end{table}

\begin{figure}
        \centering
        \resizebox{\hsize}{!}{\includegraphics{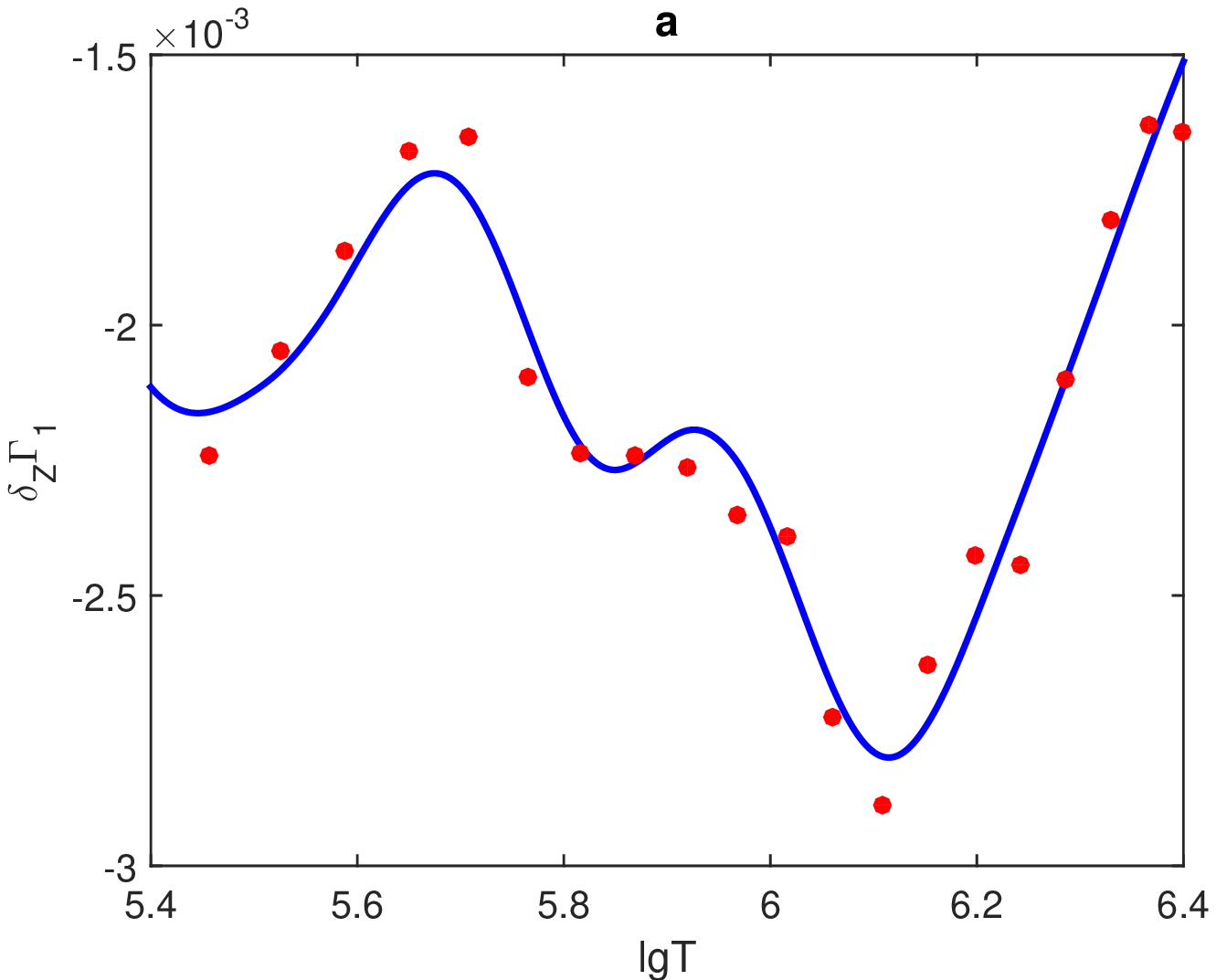}}
        \resizebox{\hsize}{!}{\includegraphics{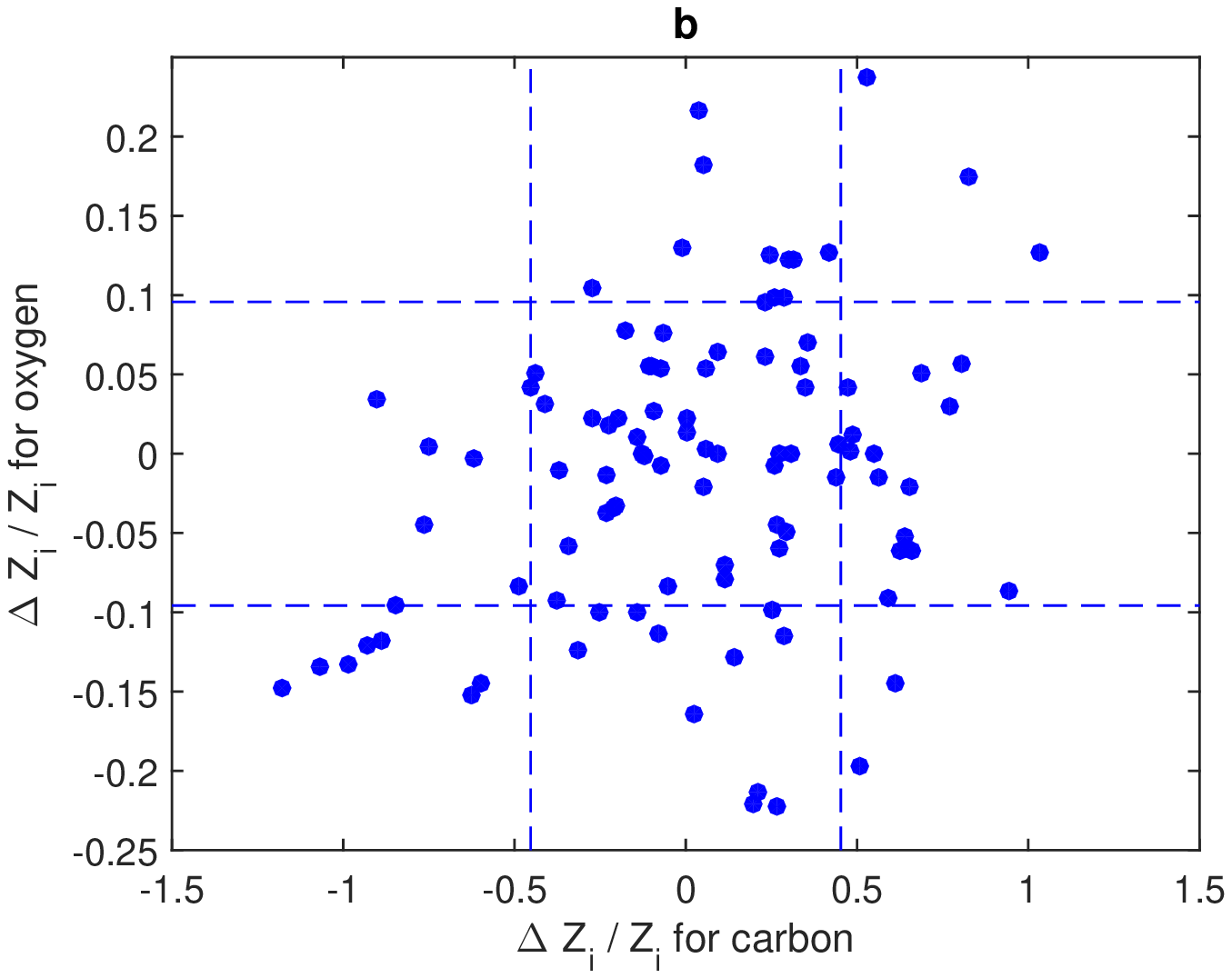}}
        \caption{(a) Precise Z contribution  $\delta_Z\Gamma_1$  from SAHA-S (blue line) and an example of random errors (red points). (b) Resulting errors of oxygen and carbon max fractions (points) in 100 experiments. Dashed lines are standard deviations normalized to main values.}
        \label{Fig_Errors}
\end{figure}

The use of the discrete version also turns out to be convenient for simulating random errors in the investigated $\Gamma_1$. We use 20 points in the $\lg T=5.4-6.4$ range. In these points, normally distributed random errors were added to the theoretical profile. Their standard deviation is $\sigma_\Gamma=10^{-4}$. The errorless $\delta_Z\Gamma_1$ from SAHA-S is presented as a blue curve in Fig.~\ref{Fig_Errors}. Red points show an example of distorted data. Only these 20 points were used for the LS decomposition.

We performed 100 experiments with different sets of 20 distorted points. The standard deviation was $\sigma_{O}=0.07\times 10^{-2}$ for oxygen and $\sigma_{C}=0.17\times 10^{-2}$ for carbon.  Fig.~\ref{Fig_Errors}b shows the relative errors of the mass fractions in every experiment. Dashed lines in Fig.~\ref{Fig_Errors}b show the standard deviations normalized to the theoretical profile. Thus, the expected error of LS-decomposition is about 10\% for
oxygen and 50\% for carbon. Estimations for other elements are not reliable.

These experiments with artificial errors show that adding random errors on the order of $\sigma _\Gamma =10^{-4}$ in the function $\Gamma_1$ will lead to a deterioration of the results compared to an error-free experiment. However, even then, the most abundant heavy element, oxygen, can be determined with a relative accuracy of about 10 percent.

\subsection{LS decomposition for $\Gamma_1(r)$  profile from standard solar model with the OPAL equation of state}
\label{Appendix_G1r}

In the final section of Appendix B, we consider an LS decomposition and estimation of heavy elements for a special illustrative example of the $\Gamma_1$  profile. This $\Gamma_1$ profile is obtained from a standard solar model calculated with the OPAL equation of state (version 2005). Here, the $\Gamma_1$ profile is considered as a function of radius, not of temperature as above. This solar model, named 772, was specifically calculated for this experiment. It was computed and calibrated identically to model 771-0001, except that the OPAL equation of state is used.

\begin{figure}
        \centering
        \resizebox{\hsize}{!}{\includegraphics{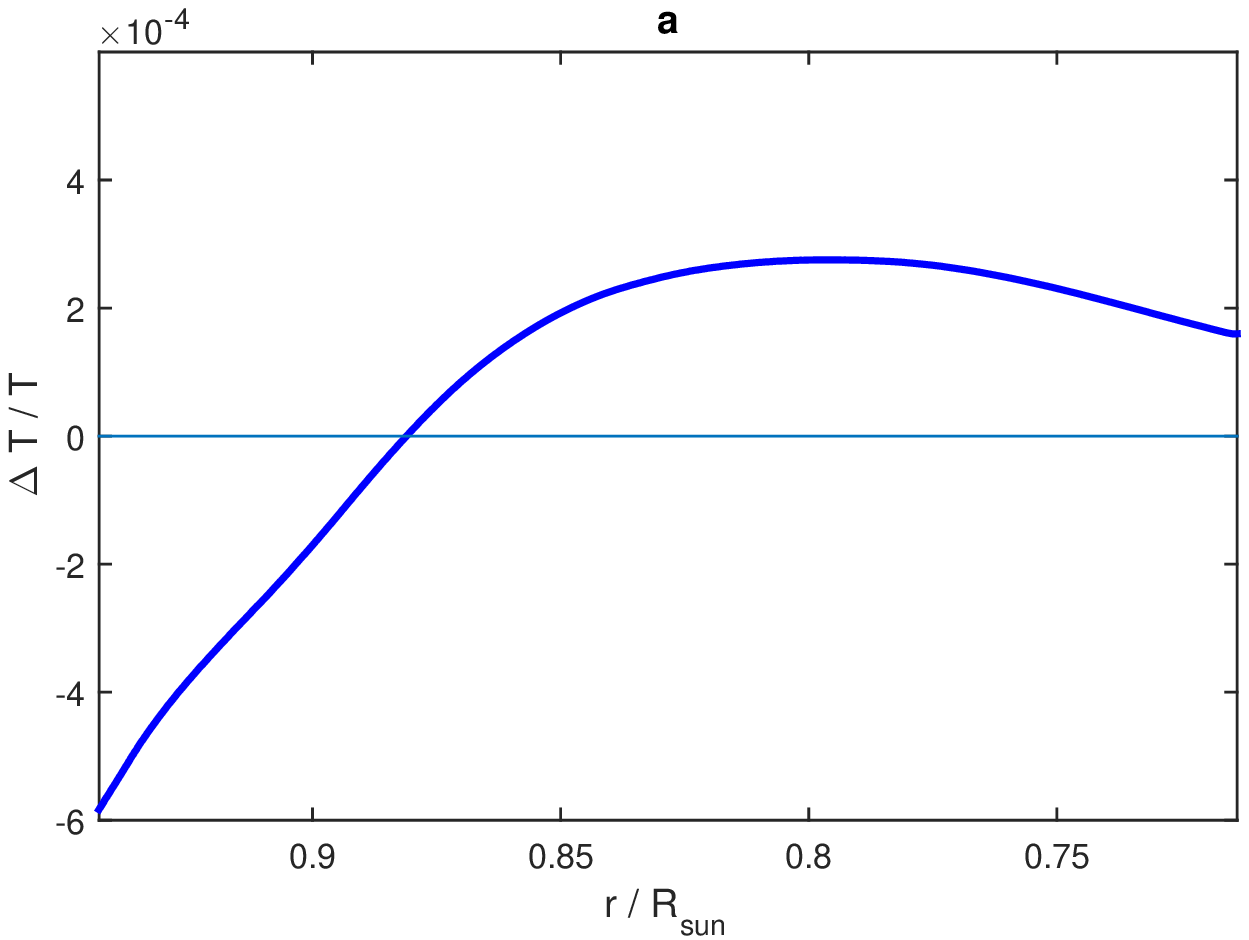}}
        \resizebox{\hsize}{!}{\includegraphics{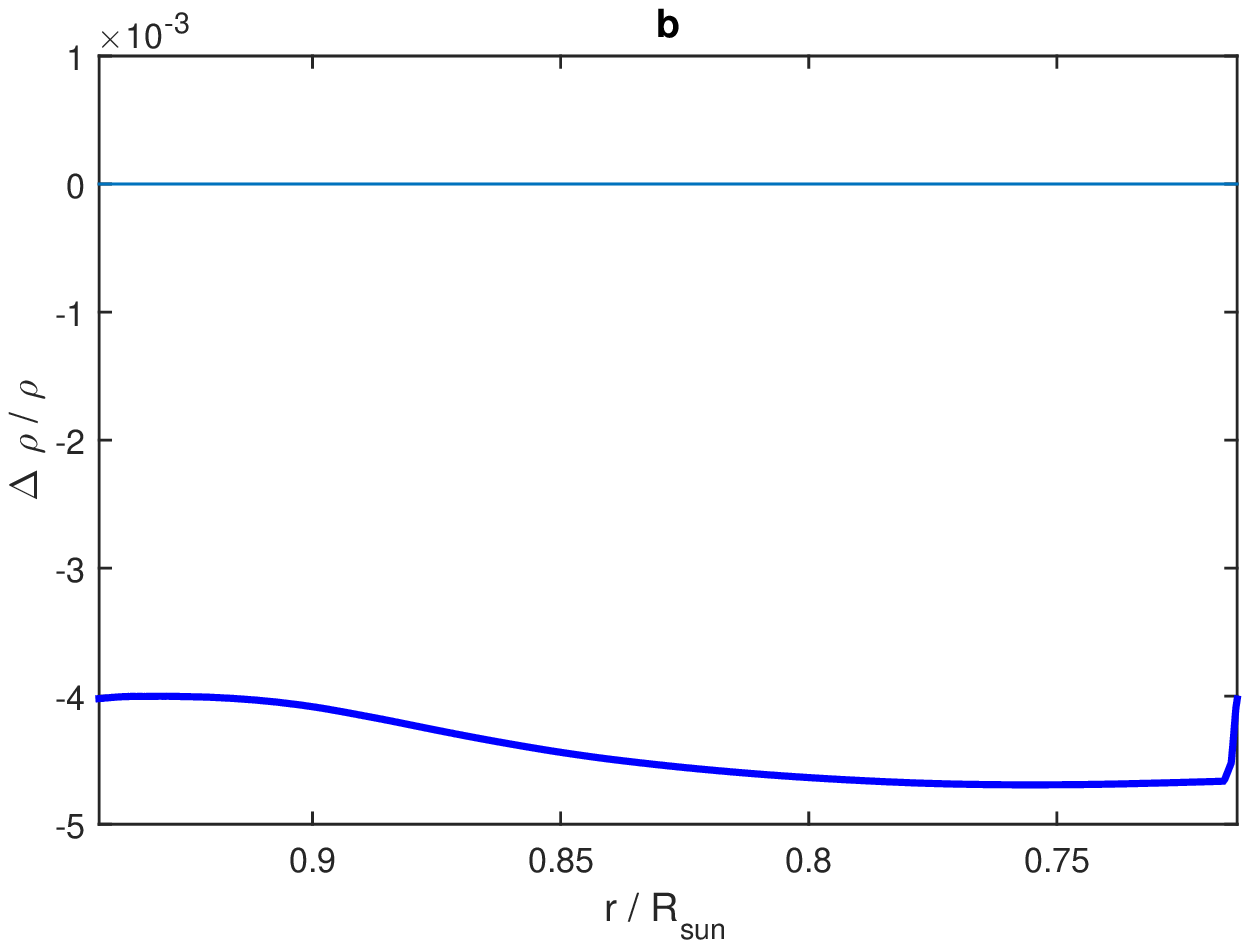}}
        \caption{Relative differences of temperatures (a) and densities (b) in models 772 and 771-0001.}
        \label{Fig_Diff_OPAL_SahaS_models}
\end{figure}

Thus, the $\Gamma_1\left( r \right)$ profiles in these models differ not only due to the different physics in the equation of state, but also due to a discrepancy in the $\left( T,\rho  \right)$ adiabats. Fig.~\ref{Fig_Diff_OPAL_SahaS_models} shows these model differences as functions of radius. In the convection zone, the temperature difference is less than 0.06\%, and it does not exceed 0.5\% in density. The density difference is mainly caused by the different adiabates of the calibrated models.

\begin{figure}
        \centering
        \resizebox{\hsize}{!}{\includegraphics{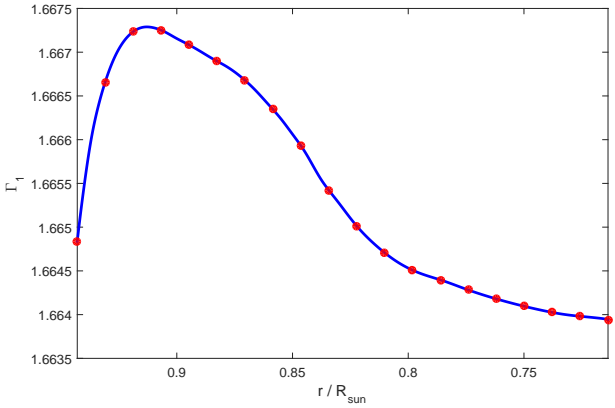}}
        \caption{Adiabatic exponent as a function of radius in model 772.
                     Red points show data used for LS decomposition.}
        \label{Fig_G1r}
\end{figure}

Following the technique proposed in Appendix~\ref{Appendix_RandomErrors}, we used 20 points uniformly distributed along the radius in the selected interval, $r/R_\odot=0.714-0.943$, which corresponds to temperatures $\lg T=5.5-6.34$. These points are plotted as red circles on Fig.~\ref{Fig_G1r}. The background $\Gamma _{1}^{HHe}$ and the basic functions $\delta_Z^i\Gamma_1$ were computed at the same points $r_j$ in the model 771-0001 and the corresponding $\left( T,\rho  \right)$ were used in the SAHA-S calculations.

As a result, the LS decomposition for the four chemical elements used in the OPAL mixture is obtained (Table~\ref{Table_G1r_fromModel}). The most represented elements (C, O, and Ne) are determined with a relative accuracy better than 10\%. The total errors are comparable with the results in Table~\ref{Table_OPALeos}.

\begin{table}
        \caption{LS decomposition for $\Gamma_1(r)$  profile from standard solar model with OPAL equation of state.}
        \centering
        \begin{tabular}{|l|c|c|}
                \hline
                Elem. & $Z_i^{\mathrm{LS}}, \times 10^{-2} $    & $\Delta Z_i/Z_i$, \% \\
                \hline  
                C &     0.3206  & -6 \\
                N &     0.1393  & 39 \\
                O &     0.9329  & -4 \\
                Ne&     0.3465  & -8 \\
                \hline
                $Z^{\mathrm{LS}}$ & \multicolumn{2}{|c|}{0.01739} \\
                \hline  
                $\eta$   & \multicolumn{2}{|c|}{$7.482\times 10^{-5}$} \\
                \hline  
                $\varepsilon$ & \multicolumn{2}{|c|}{$3.961\times 10^{-5}$} \\
                \hline  
                $\sigma$ & \multicolumn{2}{|c|}{$6.348\times 10^{-5}$} \\
                \hline
        \end{tabular}
        \label{Table_G1r_fromModel}
\end{table}

We want to note that this numerical blind-test experiment is completely illustrative. We do not draw any conclusions in addition to those already made in the article. A detailed study of individual factors and their combinations in this experiment is far beyond the scope of this paper.

\end{appendix}

\end{document}